\documentclass[a4paper,12pt]{article}
\pdfoutput=1
\usepackage{epsfig}
\usepackage{amssymb}
\usepackage{amsfonts}
\usepackage{amsmath}
\usepackage{euscript}
\usepackage{verbatim}
\usepackage{latexsym}
\usepackage{graphicx}
\usepackage{caption}
\usepackage{float}
\usepackage{xcolor}

\usepackage[hidelinks]{hyperref}

\usepackage{subcaption}

\usepackage{listings}

\usepackage{booktabs}

\newif\ifdtup

\catcode`\@=11

\@addtoreset{equation}{section}

\def\@normalsize{\@setsize\normalsize{15pt}\xiipt\@xiipt
\abovedisplayskip 14pt plus3pt minus3pt%
\belowdisplayskip \abovedisplayskip
\abovedisplayshortskip \z@ plus3pt%
\belowdisplayshortskip 7pt plus3.5pt minus0pt}

\def\small{\@setsize\small{13.6pt}\xipt\@xipt
\abovedisplayskip 13pt plus3pt minus3pt%
\belowdisplayskip \abovedisplayskip
\abovedisplayshortskip \z@ plus3pt%
\belowdisplayshortskip 7pt plus3.5pt minus0pt
\def\@listi{\parsep 4.5pt plus 2pt minus 1pt
     \itemsep \parsep
     \topsep 9pt plus 3pt minus 3pt}}

\relax

\catcode`@=12

\catcode`\@=11

\def\section{\@startsection{section}{1}{\z@}{3.5ex plus 1ex minus
   .2ex}{2.3ex plus .2ex}{\large\bf}}

\def\SymBoxes#1#2#3#4{\newdimen\un@t \un@t#3%
\raisebox{#1}{\rule{#2\un@t}{#4}\hskip-#2\un@t% lower horizontal
\@tempdimb\un@t \advance\@tempdimb by-#4\@tempcntb#2\relax%
\@whilenum{\@tempcntb>0}\do{%                         % #2 vertical lines
\rule{#4}{\un@t}\hskip\@tempdimb \advance\@tempcntb by\m@ne}%
\hskip-#2\un@t \rule[\un@t]{#2\un@t}{#4}%
\rule[\un@t]{#4}{#4}\hskip-#4%             % upper horizontal line
\rule{#4}{\un@t}}\hskip-#4}                % rightest vertical line
\begin{document}
%\begin{letter}{~}

%%%%%%Define some new commands and  macros
\newcommand{\beq}{\begin{equation}}
\newcommand{\eeq}{\end{equation}}
\newcommand{\bea}{\begin{eqnarray}}
\newcommand{\eea}{\end{eqnarray}}
\newcommand{\beas}{\begin{eqnarray*}}
\newcommand{\eeas}{\end{eqnarray*}}
\newcommand{\defi}{\stackrel{\rm def}{=}}
\newcommand{\non}{\nonumber}
\newcommand{\bquo}{\begin{quote}}
\newcommand{\enqu}{\end{quote}}
%%%%%%%%%%%%%%%%
\renewcommand{\(}{\begin{equation}}
\renewcommand{\)}{\end{equation}}
%%%%%%%%%%%%%%%%%%%%%%%%%%%%%%%%%% definitions
\def \eqn#1#2{\begin{equation}#2\label{#1}\end{equation}}

\def\e{\epsilon}
\def\IZ{{\mathbb Z}}
\def\IR{{\mathbb R}}
\def\IC{{\mathbb C}}
\def\IQ{{\mathbb Q}}
\def\de{\partial}
\def\Tr{ \hbox{\rm Tr}}
\def\H{ \hbox{\rm H}}
\def\HE{ \hbox{$\rm H^{even}$}}
\def\HO{ \hbox{$\rm H^{odd}$}}
\def\K{ \hbox{\rm K}}
\def\Im{ \hbox{\rm Im}}
\def\Ker{ \hbox{\rm Ker}}
\def\const{\hbox {\rm const.}}
\def\o{\over}
\def\im{\hbox{\rm Im}}
\def\re{\hbox{\rm Re}}
\def\bra{\langle}\def\ket{\rangle}
\def\Arg{\hbox {\rm Arg}}
\def\Re{\hbox {\rm Re}}
\def\Im{\hbox {\rm Im}}
\def\exo{\hbox {\rm exp}}
\def\diag{\hbox{\rm diag}}
\def\longvert{{\rule[-2mm]{0.1mm}{7mm}}\,}
\def\a{\alpha}
\def\dag{{}^{\dagger}}
\def\tq{{\widetilde q}}
\def\p{{}^{\prime}}
\def\W{W}
\def\N{{\cal N}}
\def\hsp{,\hspace{.7cm}}

\def\br{\nonumber}
\def\IZ{{\mathbb Z}}
\def\IR{{\mathbb R}}
\def\IC{{\mathbb C}}
\def\IQ{{\mathbb Q}}
\def\IP{{\mathbb P}}
\def \eqn#1#2{\begin{equation}#2\label{#1}\end{equation}}

\newcommand{\C}{\ensuremath{\mathbb C}}
\newcommand{\Z}{\ensuremath{\mathbb Z}}
\newcommand{\R}{\ensuremath{\mathbb R}}
\newcommand{\rp}{\ensuremath{\mathbb {RP}}}
\newcommand{\cp}{\ensuremath{\mathbb {CP}}}
\newcommand{\vac}{\ensuremath{|0\rangle}}
\newcommand{\vact}{\ensuremath{|00\rangle}                    }
\newcommand{\oc}{\ensuremath{\overline{c}}}
\newcommand{\psizero}{\psi_{0}}
\newcommand{\phizero}{\phi_{0}}
\newcommand{\hzero}{h_{0}}
\newcommand{\psiin}{\psi_{\rh}}
\newcommand{\phiin}{\phi_{\rh}}
\newcommand{\hin}{h_{\rh}}
\newcommand{\rh}{r_{h}}
\newcommand{\rb}{r_{b}}
\newcommand{\psibnd}{\psi_{0}^{b}}
\newcommand{\psibndp}{\psi_{1}^{b}}
\newcommand{\phibnd}{\phi_{0}^{b}}
\newcommand{\phibndp}{\phi_{1}^{b}}
\newcommand{\gbnd}{g_{0}^{b}}
\newcommand{\hbnd}{h_{0}^{b}}
\newcommand{\zh}{z_{h}}
\newcommand{\zb}{z_{b}}
\newcommand{\man}{\mathcal{M}}
\newcommand{\hbr}{\bar{h}}
\newcommand{\tbr}{\bar{t}}

\begin{titlepage}
%\begin{flushright} CHEP XXXXX
%ULB-TH/09-10\\
%hep-th/yymmnnn\\ \end{flushright}
%\bigskip

\def\thefootnote{\fnsymbol{footnote}}

\begin{center}
{\bf {\Large 
Building a Quantum Black Hole Microstate: \\}
\vspace{0.2cm}
{\large A Bulk Path Integral for a Heavy Virasoro Primary}
}
\end{center}

%\bigskip
\begin{center}
Chethan Krishnan$^a$\footnote{\texttt{chethan.krishnan.physics@gmail.com}}\ \ \& \ \ Rajdeep Mitra$^a$\footnote{\texttt{mitrarajdeep00@gmail.com}}
\end{center}

\renewcommand{\thefootnote}{\arabic{footnote}}

\begin{center}
%\vspace{0.2cm}

$^a$ {Center for High Energy Physics,\\
Indian Institute of Science, Bangalore 560012, India}\\

\end{center}
\vspace{-0.15in}
\noindent
\begin{center} {\bf Abstract} \end{center}
Consider level-$k$ $SL(2,\IR) \times SL(2,\IR)$ Chern-Simons theory on a semi-infinite solid cylinder, with a Wilson line in the unitary principal continuous series inserted along the time direction at the center of the spatial disc. We argue that at the $\tau=0$ cut, this Euclidean path integral prepares the bulk dual of a heavy Virasoro primary above the BTZ threshold, with Drinfel'd-Sokolov (DS) reduction playing a crucial role in converting the affine module into a Virasoro module. Semi-classically, the DS constraint turns into the DS gauge, familiar from asymptotically AdS$_3$ gravity. Together with boundary gravitons, these primaries provide a bulk construction of BTZ black hole microstates at finite $k$. In the large-$k$ WKB limit, the saddle that controls an {\em individual} primary is the BTZ black hole with a {\em singular} horizon. We study the (grand-)canonical partition function of the microstates with density of states dictated by modular invariance. The temperature and chemical potential then fix the holonomy saddle to be the Euclidean BTZ black hole with a {\em smoothly} contractible thermal cycle. A density of heavy primaries growing as $e^{2\pi\gamma QP}$ would instead lead to a defect (or excess) saddle, except at the Cardy value $\gamma=1$ fixed by modular invariance. After analytic continuation, the eternal Lorentzian black hole with a smooth horizon is therefore not just any thermal ensemble, but the ``modular ensemble" of (semi-classically singular) horizonless microstates.  We compute the thermal fluctuations in this ensemble, and reproduce the fluctuations in the area of the form expected from semi-classical general relativity. Ensembles of theories do not play a direct role in any of our discussions. We also note that these microstates, unlike fuzzballs, do {\em not} break the isometries of the black hole.

\vspace{1.6 cm}
\vfill

\end{titlepage}

\setcounter{footnote}{0}

\tableofcontents

%%%%%%%%%%%%%%%%%%%%%%%%%%%%%%%%%%%%%%%%%%%%%%%%%%%%%%%%%%%%%%%%%%%%%%%%%%%%%%%%%%%%%%%%%%%%%%
%%%%%%%%%%%%%%%%%%%%%%%%%%%%%%%%%%%%%%%%%%%%%%%%%%%%%%%%%%%%%%%%%%%%%%%%%%%%%%%%%%%%%%%%%%%%%%

\section{Introduction} 
\label{sec:Intro}

It is an oft-stated truism that a black hole is a thermal ensemble of heavy microstates. But the precise nature of these microstates in the bulk language is largely a mystery. The Gibbons-Hawking \cite{Gibbons} calculation that is the gold-standard in producing black hole thermodynamics from the bulk, is {\em not} manifestly a count of the microstates. 

It has been suggested  \cite{CKPSP}\footnote{See also motivating earlier work in \cite{Vaibhav1, Vaibhav2, Datar}.} that for generic finite temperature black holes, demanding a complete UV description of every microstate {\em in a given theory}, may not be necessary for a satisfactory understanding of black hole statistical mechanics.  The ``holographic atomic hypothesis'' advocated in \cite{CKPSP} suggests a more hopeful possibility. Just as an atomic/molecular model can capture the microscopic origin of ordinary gas thermodynamics without being too sensitive to the specific UV completion, there may exist structures that are (semi-)universal across theories of quantum gravity and are useful for describing black hole microstates. In fact, the results of \cite{CKPSP} indicate in a precise sense, that (at least in AdS$_3$/CFT$_2$) the situation is in fact {\em more} fundamental and universal for the thermodynamics of black holes, than for the thermodynamics of gases. 

The key point is that in 2D CFTs with $c > 1$, one expects a bulk description of primaries via Wilson lines in a suitable Chern-Simons theory/TQFT. This construction is usually discussed semi-classically and below the BTZ threshold, often in connection with conical defects and geodesics. But as suggested in \cite{CKPSP} and further developed in \cite{Vishal, Vishal2} and in the present paper, it is a much more general structure. It implies that the nature of microstates is universal in AdS$_3$/CFT$_2$ across theories: what differs from theory-to-theory is the {\em specific} set of discrete primaries and their ``interactions" as controlled by the OPE coefficients. In other words, the non-perturbative picture of the bulk is that of a TQFT skeleton \cite{Vishal2} together with its sewing-compatible (discrete) bootstrap data.

While the ingredients are quite well-known, this perspective may not be -- so let us explain this in a bit more detail. What we are suggesting is that we may already have a complete description of non-perturbative bulk AdS$_3$ quantum gravity, via {\em bulk} bootstrap in terms of a non-compact TQFT\footnote{We emphasize that {\em explicitly} solving for sewing compatible bootstrap data is a tall order. So while we do believe this is a powerful re-frame, it is not the same things as the {\em ability} to solve explicit theories. This picture should be compared to that in rational CFTs, where the bulk (compact) TQFTs are completely solvable and the number of primaries (of some extended chiral algebra) is finite. }. A two-dimensional CFT naturally separates into universal
kinematic data and theory-specific data.  Virasoro representation
theory fixes the characters, conformal blocks, and the kernels that relate
different sewing channels.  A particular CFT selects which primary
representations actually occur and specifies their OPE coefficients.  In the
bulk, the first layer has a natural realization as a non-compact topological
skeleton, closely related to Teichm\"uller or Virasoro TQFT
\cite{Kashaev, Teschner1, Teschner2, Teschner, Andersen, CEZ1, CEZ2, CKPSP, Vishal, Vishal2}.  Wilson lines carry representation
labels, Wilson networks compute conformal blocks, and changes of bulk
decomposition implement the same sewing moves that underlie the CFT
bootstrap on all surfaces with insertions.  From this perspective, the Wilson line is not merely a useful
probe inserted into an already existing geometry.  It is the natural
topological carrier of the quantum numbers of a primary state.

The Chern-Simons/TQFT formulation allows the Euclidean horizon to be replaced by an inner boundary carrying holonomy data, and the modular decomposition of the
BTZ character can then be interpreted as a sum over bulk states which are not individually described by the smooth Euclidean black-hole geometry \cite{CKPSP}.  The smooth horizon partition function appears only after the appropriate spectral sum, as was missing from the Gibbons-Hawking calculation. In the Cardy regime \cite{Cardy}, the discrete sum over explicit microstates in a specific holographic CFT is extremely well-approximated by the continuum sum over the $S$-kernel density of states \cite{CKPSP}. This can be viewed as the ``smooth horizon" approximation.

These ideas naturally suggest a concrete bulk microscopic interpretation of heavy
CFT primaries.  Once an individual compact CFT contains a primary with given
$(h,\bar h)$, the topological skeleton should contain a corresponding bulk
state labeled by the associated Wilson line.  The skeleton by itself does
not determine which such primaries occur in the exact theory, nor their
multiplicities.  Those are part of the theory-specific CFT data.  What it
does provide is a universal bulk realization of the state associated with
each allowed primary.  Below the BTZ threshold this statement is familiar
from conical defects, which are associated with elliptic holonomy and
discrete-series representations.  Above threshold the analogous object has
hyperbolic holonomy.  This points to Wilson lines in the principal continuous
series as the natural bulk representatives of heavy Virasoro primaries.

One purpose of this paper is to make this statement explicit at the level of
the bulk state.  We consider $SL(2,\IR)\times SL(2,\IR)$
Chern-Simons theory on a semi-infinite solid cylinder and insert a Wilson
line along the Euclidean time direction at the center of the spatial disc.
The Euclidean path integral then prepares a state on the $\tau=0$ cut.  The
state first lives in an affine current-algebra module.  The physical
gravitational state is obtained only after imposing the reduction associated
with the asymptotically AdS$_3$ boundary conditions.  In the language used
below, this is a quantum Drinfel'd-Sokolov reduction.  The result is a
Virasoro primary together with its boundary-graviton descendants.  Thus the
Wilson-line state provides a bridge between the finite-$c$ TQFT skeleton and
the more familiar bulk description in which a semiclassical bulk
geometry can be extracted from the gauge field.

A fixed heavy
primary is a state with a definite Wilson-line label and definite
hyperbolic holonomy (but with no a priori thermal periodicity).  In the large-$k$ limit its saddle is the BTZ gauge field, but
with the Wilson-line source.  In the metric language, a semi-classical primary is therefore naturally interpreted as having a singular horizon, a place where ``spacetime ends'' in the sense of geodesic incompleteness\footnote{Note that geodesics are probes that only make sense on coherent heavy backgrounds, so on a microstate, we are using it only as an intuitive tool rather than as a physical probe.}. Smoothness emerges after summing over heavy primary sectors. In the Cardy regime,
modular invariance fixes the universal continuum approximated
density through the vacuum row of the modular $S$-kernel.  Extremizing this
modularly weighted partition sum over the Wilson-line label selects precisely the
holonomy for which the Euclidean thermal cycle becomes smoothly
contractible.  In other words, the Gibbons-Hawking regularity condition is an
emergent saddle-point statement about the spectrum rather than a microscopic
condition imposed on every state.

The resulting picture is the one we will develop throughout the paper.  The
fundamental finite-$c$ objects are quantum states, while a smooth black-hole geometry is a collective semiclassical
description of an appropriate set of such states.  The exact spectrum of an
individual CFT remains discrete.  The modular kernel supplies its universal
continuum description in the regime in which Cardy coarse graining is
valid.  No ensemble of theories is required.  What emerges is instead a
``modular ensemble'' of states within a theory.  In this language, the
holographic atomic hypothesis comes into its own: the relevant atoms are
heavy Wilson-line sectors, and the smooth BTZ horizon is a collective
property that emerges from them.  Our first task is therefore to explain how the Wilson-line state is prepared and how its affine
current-algebra description is converted into the physical Virasoro
microstate.

\subsection{Strategy: Quantize-then-Reduce}\label{sec:strategy}
 
The central technical ingredient in this paper is a two-step construction.
A Wilson line along the Euclidean time axis first prepares a state in
an \emph{unreduced} affine $\widehat{\mathfrak{sl}}(2,\IR)_k$ module on
the punctured spatial disc. The one-unit spectrally flowed
Drinfel'd-Sokolov (DS) reduction appropriate to the Brown-Henneaux boundary
conditions on the cylinder \cite{CHvD,BrownHenneaux} is then used to pass from
this affine module to the physical Virasoro module. Since the unreduced quantum $SL(2,\IR)$
Chern-Simons theory is subtle\footnote{Note that this correlates with the fact that pure 3D Einstein gravity is likely not the best starting point for a unitary quantum gravity.} -- the gauge group is
non-compact and a satisfactory quantization is not available to our knowledge -- we first orient the reader on how this strategy relates to previous
approaches, in which the reduction appears in different guises.
 
In the Wilson-line computations of Virasoro blocks
\cite{Besken:2016ooo,Fitzpatrick:2016mtp,Besken:2017fsj,Hikida:2018eih,Besken:2018zro},
the reduction enters as a renormalization condition rather than as an
operation on a Hilbert space.  Matrix elements of the
$\mathfrak{sl}(2)$ Wilson line are divergent at the quantum level, and
the ambiguities are fixed by demanding consistency with \emph{Virasoro}
(rather than current-algebra) Ward identities. The resulting anomalous
dimension shifts the naive weight to its DS/Liouville value. The
choice of which Ward identities define the quantum theory can be viewed as the place where 
reduction enters\footnote{While we believe this is natural, it should be emphasized that this is best viewed as an {\em interpretation} of the calculation.}, imposed order by order in $1/c$, even though the constraint itself is
never exhibited acting on a module.  In the coadjoint-orbit approach
\cite{Witten:1987ty,Alekseev:1988ce,Raeymaekers:2014kea,CJ1}, the
reduction is instead performed classically, before quantization: one
starts from the already-reduced phase space -- Virasoro coadjoint
orbits, equivalently Ba\~nados geometries with Brown-Henneaux boundary
conditions -- and quantizes that.  This route never encounters the
unreduced quantum theory at all.  Finally, in the Virasoro TQFT of
\cite{CEZ1, CEZ2}, the reduction is built into the
very definition of the theory: the Hilbert spaces are \emph{defined} to
be spaces of Virasoro conformal blocks, so that the outcome of the
reduction is taken as an axiom, and defects enter directly as Virasoro
primary labels.  In none of these approaches is the unreduced affine
puncture sector on the disc written down.  Our attitude is that the affine module is
nevertheless useful as an intermediate ``upstairs" description:
we never rely on a full quantization of $SL(2,\IR)$ Chern-Simons
theory, only on the fixed Wilson-line sector, which we do eventually reduce. 

The utility of this ``upstairs" object is that it allows us to make connections with semi-classical {\em bulk} solutions in the large-$k$ limit. 
We find that this gives clarity on multiple aspects of the physics: specifically in matters of dynamics, semi-classics and bulk geometry.  First, the
construction is Hartle-Hawking-like. An explicit Euclidean bulk path
integral with a Wilson-line source prepares the unreduced affine state on the
$\tau=0$ cut (and the flowed Brown-Henneaux reduction then gives the physical
Virasoro state).  The reduce-then-quantize route hands one
the Hilbert space, but has no counterpart of this bulk preparation.
Second, the semi-classical discussion of
Sections~\ref{sec:semiclassical} and~\ref{sec:thermal} -- the WKB limit
of the wavefunctional, the saddle-point connection, and the resulting
singular-versus-smooth horizon geometries -- is a statement about bulk
saddles of the Chern-Simons theory and allow a natural bulk metric connection.  The coadjoint-orbit
description is intrinsically a boundary theory, with no bulk interior
whose smoothness can be interrogated. Keeping the connection as the
basic variable is what allows us to ask questions about the horizon.  Third, the bulk mechanism that distinguishes the sub- and
super-threshold sectors -- elliptic holonomy and conical defects versus hyperbolic holonomy and black hole microstates -- is more intuitive upstairs, whereas
downstairs they are all encoded as normal orbits. 

Our approach also provides us a comparison\footnote{Or perhaps a consistency check.} between quantize-then-reduce and reduce-then-quantize. Let us explain this. For the sub-threshold
(discrete-series) sector of Section~\ref{sec:microstates}, the affine
module is of lowest-weight type\footnote{A lowest-weight module is a
highest-weight module for the opposite Borel subalgebra, so we use the two terms somewhat
interchangeably.}. For the unflowed plane
reduction this lies within the standard highest-weight quantum DS framework
\cite{FKW, dBT, BersOoguri,FeiginFrenkel,ArakawaVanishing,ArakawaReps}.
The gravitational problem considered here, however, is the one-unit
spectrally flowed cylinder reduction, so we do not invoke the standard
unflowed reduction theorems as an all-level proof. The super-threshold
(continuous-series) sector of Section~\ref{sec:super-threshold}
has the additional feature that the affine module is relaxed\footnote{The terminology on affine representations is a bit baroque in the mathematics literature. What we call a relaxed module induced from the principal continuous series seems to be called (extremely confusingly!) a  ``relaxed highest weight" representation in some places. We will not follow that nomenclature. As a mnemonic: we will have relaxed above threshold and highest/lowest-weight below.},
with no highest- or lowest-weight vector, and hence lies still further outside
the usual extremal-weight setting.  For the principal series the level-zero
mechanism can nevertheless be analyzed directly: at fixed $s$ and fixed
$\epsilon$, the cylinder constraint selects a single Virasoro primary class,
as spelled out in Appendix~\ref{app:DS-reduction}.  The  positive-level
BRST cohomology of the flowed discrete and relaxed modules will be discussed in
\cite{Future}\footnote{Spectrally flowed DS reductions are discussed from a mathematical perspective in \cite{Dhillon}.}.  The all-level structure -- for generic
super-threshold momentum, a single non-degenerate Verma module built on a
primary with $h>\frac{c-1}{24}$ -- is also what reduce-then-quantize produces
from the quantization of the hyperbolic ${\rm Diff}(S^1)/S^1$ coadjoint orbit
\cite{Witten:1987ty,Alekseev:1988ce,CJ1, CKPSP}.  We therefore view the two
routes as complementary: the orbit quantization corroborates the
reduced Hilbert space, while the Wilson-line construction supplies
what the orbit description cannot -- the bulk preparation of the
microstate and its saddle-point geometry.

\subsection{Microstates and the Nature of the Smooth Horizon}

In the second half of the paper (sections~\ref{sec:semiclassical} and~\ref{sec:thermal}) we discuss the relationship between individual black-hole microstates and a smooth
black-hole geometry.  A heavy primary defines a perfectly good quantum
state, represented in the bulk by a Wilson line whose conjugacy class
fixes the BTZ parameters.  In the large-$k$ limit, the corresponding
geometry is locally BTZ and contains a finite-perimeter
would-be horizon.  However, the state-preparation path integral does
not impose the thermal identification required for smoothness of the
Euclidean horizon.  The semiclassical (i.e., large-$k$) geometry of an individual
microstate therefore ends at a singular finite-perimeter surface,
rather than extending through a smooth Kruskal horizon. Note that this singularity of the microstate is a semi-classical artifact -- at finite-$k$ it is a well-defined quantum state.

The smooth geometry emerges only after summing over heavy states.
In the Cardy regime, modular invariance fixes the asymptotic density of
primaries through the vacuum row of the Virasoro $S$-kernel.  Since the
integration variable is also the Wilson-line label, the thermal sum is
directly a sum over bulk holonomies.  At large $k$, the competition
between the Cardy density and the Boltzmann weight gives
\beq
  \lambda_L=\frac{2\pi}{\beta_L}\,,
  \qquad
  \lambda_R=\frac{2\pi}{\beta_R}\,,
\eeq
which are precisely the conditions for smoothness of the Euclidean BTZ
horizon.  The Gibbons--Hawking regularity condition is therefore not a
property imposed on each microstate, it appears as the saddle-point
condition of the thermal sum.

This also shows why {\em any} thermal ensemble in itself is not sufficient.  If the
Cardy growth exponent of the primary density is rescaled by a factor $\gamma$,
the saddle instead gives
\beq
  \lambda_{L,R}\beta_{L,R}=2\pi\gamma\,.
\eeq
The resulting geometry has a conical defect or excess at the horizon,
and simultaneously
  $S=\gamma\,\frac{A}{4G_N}$. Thus smoothness and the Bekenstein--Hawking relation are tied to the
same microscopic input: the modularly determined growth of states.
Modular invariance supplies
the spectral weighting required for the thermal state to have a smooth
semiclassical horizon. In Lorentzian signature,
although a thermofield-double purification can be written for any
thermal density matrix, a smooth Einstein--Rosen bridge and Kruskal
extension naturally require such a \emph{modular ensemble}.

The entropy itself has a similarly direct interpretation.  The
$S$-kernel gives the Cardy growth of heavy Virasoro primaries, while
boundary-graviton descendants provide the remaining contribution, and
together they reproduce the full Bekenstein-Hawking entropy \cite{CKPSP}. The horizon area is therefore associated with the dominant holonomy in
a statistical ensemble of Wilson-line states.

Finally, thermal fluctuations around the saddle are fluctuations of the same
Wilson-line label, or equivalently of the horizon holonomy.  They give
\beq
  \operatorname{Var}(S)=S\,,
  \qquad
  \operatorname{Var}(A)=4G_N A\,,
  \qquad
  \frac{\Delta A}{A}\sim S^{-1/2}\,.
\eeq
The smooth horizon is consequently the sharply peaked,
large-entropy limit of this distribution.

The resulting picture is that heavy primaries furnish the microscopic
BTZ states, but smoothness of the horizon is a collective
property of the thermal ensemble with a modular density of states.

\subsection{Comments}

Let us conclude the Introduction with some technical and conceptual observations.
\begin{itemize}
\item The Drinfel'd-Sokolov reduction we do in this paper naturally leads us to a central charge for the vacuum AdS$_3$ that is consistent with the ``regulate-and-quotient" prescription of \cite{CKPSP}. This strengthens the case\footnote{See \cite{Benjamin} for an earlier calculation that naturally enables this interpretation.} that the classical central charge $C$ of the orbit and the quantum central charge are related by $c=C+1$ across orbits. This is in contrast to the special treatment for the exceptional orbit that is sometimes suggested, with $c=C+13$. We discuss this in more detail in Section \ref{sec:BRST}.
\item In the TQFT skeleton language, the construction of the Virasoro primary (and the associated Verma module) from the Wilson line is essentially tautological. This is because the Wilson labels there are automatically those of Virasoro and not affine. But the connection with semi-classical gravity and the metric description is clearer in the Chern-Simons language. Apart from allowing us to connect with the familiar BTZ language and the role of smooth and unsmooth horizons, the Chern-Simons/DS reduction also clarifies the role of Brown-Henneaux boundary conditions from a quantum-first perspective.
\item A feature of the construction that was noted in \cite{CKPSP, PradiptaVishal} is that hyperbolic holonomy Wilson line sources lead to metric configurations where the source has a ``size", equal to the horizon size of the associated black hole. We do not believe there is any contradiction here: the Wilson line is placed at the origin of the coordinate disc, and is a topological object, while size is a metric notion. (Hyperbolic) holonomy in one language translates to horizon size in the other.  This simple yet remarkable observation does not seem to be as widely appreciated as it should be.
\item Another fact, which was again mentioned in \cite{CKPSP} is that these microstates do not break the isometries of the BTZ black hole. This is satisfying because in many BPS index computations, it is known that the microstates that contribute to a black hole's index preserve its isometries \cite{Sen1,Sen2,Sen3}. This is sometimes used as an argument against fuzzball constructions \cite{Fuzz1,Fuzz2} that break the symmetries of the horizon \cite{Sen2,Polchinski}.
%\item It was suggested in \cite{CKPSP, Vishal} that these observations are not limited to AdS$_3$/CFT$_2$ and should generalize to higher dimensions. We have been able to make progress on this question, by identifying the higher dimensional TQFT analogue to be BF theory (with the AdS isometry group as the gauge group) \cite{CK}. There are two key differences. Firstly, the presence of an independent stress tensor in the CFT means that the bootstrapped spectrum necessarily contains an independent spin-2 primary that is a conserved current. This translates to a condition that generalizes the so-called ``simplicity condition" familiar from loop quantum gravity and spin-foam models: simplicity is what causes BF theory to reduce to Einstein gravity. The generalized simplicity condition we obtain from the presence of the stress tensor in the spectrum, can be viewed as allowing Einstein gravity together with higher derivative corrections. Secondly, BF theory allows a natural generalization of modular invariance in the bulk as a Fourier transform between the B and F charges (which are canonically conjugate in the TQFT). This is a direct generalization of the observation in \cite{CKPSP} that in AdS$_3$/CFT$_2$ boundary modular transformation is implemented as a Fourier transform in the Chern-Simons/TQFT. 

\item The modular ensemble perspective allows us to understand why the smooth horizon is such a persistent and compelling feature of black hole physics, despite the microstates (in the semi-classical limit) being singular at the horizon. The key point is that the Cardy regime is a well-defined regime even at finite-$c$. So vacuum dominance in the dual channel and utility of the $S$-kernel remain intact as valid approximations even away from the semi-classical limit. Some simple calculations that quantify this point, will be presented in a companion paper \cite{Apoorva}. 

\item  We mentioned earlier in a footnote that geodesics are not the best tools to characterize semi-classical microstates, because they are naturally defined on coherent heavy backgrounds. More fundamentally, we believe that while we can write down a metric that is translated from the Chern-Simons field in this (large-$k$) limit, the \emph{physical} significance of the metric language itself is questionable. The semi-classical microstate is perhaps best left as a Chern-Simons saddle with a source. In our discussions in Section \ref{sec:semiclassical}, such caveats about the {\em utility} of individual semi-classical microstates should be kept in mind. While it is instructive to study such objects for some purposes, the dynamics of a light probe interacting with the heavy microstate is a detailed question that also involves OPE coefficients (and ETH, and such). Our eigen-microstates enable such questions to be meaningfully asked, but we leave them for the future. Given our findings, it is a natural possibility that a smooth interior will emerge from the dynamics/interaction of the probe (say, a light primary) with a single heavy microstate \cite{Jared1, Jared2}, when the probe has only coarse-grained resolution: this is when we expect that the heavy microstate can be replaced by the ensemble of horizonless microstates that act like a smooth horizon. One suspects that this, is what underlies infall.

\end{itemize}

\section{Conical Defect Microstates and Sub-Threshold Primaries}\label{sec:microstates}

Before turning to the black-hole regime, it is useful to recall the sub-threshold version of the construction.  A conical defect in AdS$_3$ is characterized by an elliptic holonomy around the particle worldline.  In Chern-Simons language this holonomy is naturally implemented by a Wilson line piercing the spatial disc.  The Wilson
line label specifies a sector of the bulk Hilbert space, and the Euclidean path integral on a half-cylinder prepares the corresponding
representation-valued affine state.  Keeping the open Wilson-line index free retains the whole horizontal\footnote{Here and below,
``horizontal'' refers to the ordinary $\mathfrak{sl}(2,\IR)$ subalgebra
generated by the affine zero modes $J^a_0$.  A horizontal
representation is therefore the representation carried by the
level-zero states before the nonzero current modes act.} multiplet, see Appendix \ref{sec:compact-prototype}.
Contracting it with an endpoint vector chooses an upstairs (affine)
representative.  This choice
will not survive as a multiplicity of gravitational primaries after
the Drinfel'd-Sokolov reduction.

The conical-defect
case is the simpler sub-threshold analogue of the construction developed
in Section~\ref{sec:super-threshold}.  It illustrates three structural
features that will persist above the BTZ threshold:
\begin{enumerate}
\item a Wilson line on the time axis turns the spatial disc into a
  punctured disc;
\item the puncture label fixes the conjugacy class of the linking
  spatial holonomy;
\item after imposing the gravitational boundary conditions -- the
  Drinfel'd-Sokolov reduction described in
  Section~\ref{sec:microstate-tower} -- the horizontal affine data
  reduce to a Virasoro primary and, at positive level, the expected
  Virasoro descendant tower.  We will present the general structure here, but the detailed all-level BRST cohomology is left for \cite{Future}.
\end{enumerate}
We use the phrase ``fixed conical defect''
to mean the primary label, or holonomy conjugacy
class.  Descendants are boundary-graviton dressings of this primary
sector.  They carry energy:
Virasoro descendants have $L_0=h+N$ and
$\bar L_0=\bar h+\bar N$, but they share the same 
primary/holonomy label.

\subsection{AdS$_3$ Gravity from Chern-Simons Theory}\label{sec:CS-review}

We begin by fixing conventions.  Three-dimensional Einstein gravity
with negative cosmological constant $\Lambda=-1/\ell^2$ can be written
as a difference of two Chern-Simons theories
\cite{Witten:1988hc,Achucarro:1987vz}. At the Lie-algebra level the
gauge symmetry is $\mathfrak{sl}(2,\IR)_L\oplus
\mathfrak{sl}(2,\IR)_R$\footnote{Throughout, our Euclidean continuation is the coordinate Wick
rotation $t=-i\tau$ of the Lorentzian Chern-Simons description, with
the noncompact gauge algebra and Wilson-line representation labels
kept fixed.
This follows Section 2 of \cite{Ammon} and Appendix~B of
\cite{CKPSP}. At the level of the group, we use the universal cover,
so the discrete-series label $j$ need not be quantized.}.  The two gauge fields are
\beq\label{eq:CS-connections}
  A \;=\; \Bigl(\omega^a+\frac{e^a}{\ell}\Bigr)J_a\,,
  \qquad
  \widetilde A \;=\;
  \Bigl(\omega^a-\frac{e^a}{\ell}\Bigr)J_a\, .
\eeq
Here $e^a$ is the dreibein, $\omega^a$ is the spin connection, and
$J_a$ are generators of $\mathfrak{sl}(2,\IR)$ satisfying
\beq\label{eq:sl2-comm}
  [J_a,J_b] \;=\; \epsilon_{abc}\,\eta^{cd}J_d\,,
  \qquad
  \eta^{ab}=\diag(-1,+1,+1)\, .
\eeq
$J_a$ denotes a basis of the finite-dimensional classical Lie
algebra.  In the Chern-Simons/gravity formulas, $\Tr$ denotes the
fundamental-representation trace.  For the Cartan--Weyl basis used
below we choose
\[
 H=\frac12
 \begin{pmatrix}1&0\\0&-1\end{pmatrix},
 \qquad
  E=
 \begin{pmatrix}0&1\\0&0\end{pmatrix},
 \qquad
  F=
 \begin{pmatrix}0&0\\-1&0\end{pmatrix},
\]
so that
\[
 [H,E]=E,\qquad [H,F]=-F,\qquad [E,F]=-2H,
\]
and
\[
 2\Tr(HH)=1,\qquad 2\Tr(EF)=-2.
\]
The affine current algebra below uses the same Lie-algebra basis and
commutation relations, $(J^3, J^+,J^-) \equiv (H,E,F)$. But there, the invariant form will be the opposite: 
$\kappa(X,Y)\equiv -2\Tr(XY)$, with 
$\kappa^{33}=-1$ and $\kappa^{+-}=2$.
We reserve $J^a_n$ for the corresponding affine-current modes.
The action, which is equal to the Einstein action up to boundary terms, is
\beq\label{eq:CS-action}
  S \;=\; S_{\rm CS}[A]-S_{\rm CS}[\widetilde A]\,,
  \qquad
  S_{\rm CS}[A]
  \;=\;
  \frac{k}{4\pi}\int_{\mathcal M}
  \Tr\!\left(A\wedge dA+\frac{2}{3}A\wedge A\wedge A\right),
\eeq
with
\beq\label{eq:level}
  k \;=\; \frac{\ell}{4G_N}\, .
\eeq
At leading semiclassical order the Brown-Henneaux central charge is
$c=6k$.  We will use this leading relation for semiclassical
matching.

\subsection{Conical Defects and Elliptic Holonomy}\label{sec:conical-defects}

A spinless massive particle at the centre of AdS$_3$ produces a
conical defect when its mass lies below the BTZ threshold.  A convenient
Euclidean form of the metric is
\beq\label{eq:conical-metric}
  ds^2 \;=\; (r^2+\alpha^2)\,d\tau^2
  +\frac{dr^2}{r^2+\alpha^2}
  +r^2\,d\phi^2\,,
  \qquad
  \phi\sim \phi+2\pi\, ,
\eeq
where $0<\alpha\leq 1$.  The deficit angle is
$2\pi(1-\alpha)$.  The endpoint $\alpha=1$ is global AdS$_3$, while
$\alpha\to0^+$ approaches the BTZ threshold from below.  One definition of the ADM mass
is
\[
  8G_N M \;=\; - \alpha^2\, ,
\]
and the corresponding left and right conformal weights of the
spinless primary are
\beq\label{eq:h-alpha}
  h_L \;=\; h_R \;=\;
  \frac{c}{24}\bigl(1-\alpha^2\bigr)\, .
\eeq
Thus $h=0$ for global AdS$_3$ and $h\to c/24$ at the BTZ threshold.

The same distinction is expressed gauge-theoretically by the holonomy
around a small spatial circle linking the particle worldline\footnote{We will often use the phrase ``particle worldline" and ``Wilson line" interchangeably. In the spinless elliptic (conical defect) case, in a suitable semi-classical limit, the two are the same. But the geodesic limit will not be our concern in this paper.}.  For a
conical defect this holonomy lies in an elliptic conjugacy class.  In the fundamental representation one may write
\beq\label{eq:elliptic-holonomy}
  {\rm Hol}(A)
  \;\sim\;
  \begin{pmatrix}
    e^{i\pi\alpha} & 0 \\
    0 & e^{-i\pi\alpha}
  \end{pmatrix},
  \qquad
  \Tr_{\bf 2}\,{\rm Hol}(A)=2\cos(\pi\alpha)\, .
\eeq
The matrix in \eqref{eq:elliptic-holonomy} should be understood as a
complex-diagonal representative of the real elliptic conjugacy class of
$SL(2,\IR)$.  The corresponding statement for the right-moving
connection is analogous (including for spinning conical defects).

\subsection{Wilson Lines as Sources}\label{sec:Wilson-line}

A massive particle in Chern-Simons theory is naturally described by a
Wilson line supported on its worldline.  Let $\gamma$ be the line
$\{0\}\times \IR_\tau$ at the centre of the spatial disc.  A Wilson line
in a representation $R$ of $SL(2,\IR)$ may be denoted schematically as
\beq\label{eq:Wilson-def}
  W_R(\gamma)
  \;=\;
  \Tr_R\,{\cal P}\exp\!\left(\int_\gamma A\right),
\eeq
or, more precisely for an open line ending on a time slice, by the
corresponding worldline path integral with endpoints in the
representation space of $R$.

The physical datum carried by the line is the conjugacy class of the
holonomy of the bulk connection around a small circle linking the line.
Equivalently, at fixed Euclidean time the spatial slice is
$D^2\setminus\{0\}$, and the Wilson line labels the puncture at the
origin.  In the sub-threshold case this linking holonomy is elliptic, in the super-threshold case it will be hyperbolic.

The coadjoint-orbit description is a convenient way to describe this
\cite{Alekseev:1988vj,Alekseev:1994,Basile:2023coadjoint}. Even though
we will not need it much, the intuition is useful and so we provide a
quick summary of some facts. The idea here is that one can write the Wilson line
above \eqref{eq:Wilson-def} using an auxiliary path integral over a
group-valued field on the worldline. Concretely, one introduces a field
$U(s) \in SL(2,\IR)$ on the worldline with first-order action
$\int ds \, {\rm Tr}\!\left(\mu\, U^{-1} D_s U\right)$, where $\mu$ is a
fixed Lie-algebra element encoding the mass and spin of the particle,
and $D_s$ is the covariant derivative along the worldline. Because the
action is first order, it defines a phase space rather than a
configuration space: the physical phase space variable is a point on the ``co-adjoint orbit" of $\mu$:
\[
  Q^{\rm orb} \;=\; U \mu\, U^{-1} \;=\; q^a J_a \;\in\; \mathfrak{sl}(2,\IR)\, ,
\]
where the coadjoint charge has been identified with a Lie-algebra
element using the invariant bilinear form appearing in the
Chern-Simons action.\footnote{This trace pairing fixes the
normalization of the classical source and holonomy.  The operator
Casimir below is written in a Cartan--Weyl convention adapted to the
quantum representation. We therefore keep the classical orbit radius
and the quantum label $j$ distinct except in their semiclassical
matching.}  The charge lives on the set of all conjugates of $\mu$ --
which is the coadjoint orbit ${\cal O}_\mu$ of $SL(2,\IR)$. The internal degree of
freedom of the particle is thus a classical mechanical system on
${\cal O}_\mu$, coupled to the pullback of the Chern-Simons connection
through the term linear in $A_s$. Quantizing this orbit -- which can be implemented by the path integral over $U$ -- gives a representation
$R$. The conjugacy class of $Q^{\rm orb}$ is fixed by $\mu$, and upon quantization
its components $q^a$ are promoted to the generators of $R$.
The equation of motion in the presence of the line source is
\beq\label{eq:CS-eom-source}
  F_{ij}(x)
  \;=\;
  \frac{2\pi}{k}\,q^a\,\delta^{(2)}(x)\,
  \epsilon_{ij}\,J_a\, ,
\eeq
so the connection is flat away from the puncture.  On the punctured
disc one may write, up to conjugacy,
\beq\label{eq:flat-Aphi}
  A_\phi \;=\; \frac{1}{k}\,Q^{\rm orb} \;\sim\; \frac{1}{k}\,\mu\, ,
  \qquad
  {\rm Hol}(A)
  \;=\;
  \exp\!\left(2\pi A_\phi\right)
  \;=\;
  \exp\!\left(\frac{2\pi \mu}{k}\right).
\eeq
Thus the orbit label $\mu$ -- equivalently, the representation data of
the Wilson line -- determines the conjugacy class of the linking
holonomy.  

\subsection{The Unreduced Conical-Defect Sector}\label{sec:SL2R-adaptation}

We now discuss the Wilson-line construction. The
Wilson line first prepares a puncture sector of the unreduced
$\widehat{\mathfrak{sl}}(2,\IR)_k$ current algebra.  Brown-Henneaux
boundary conditions are imposed by Drinfel'd-Sokolov
(DS) reduction.  The Virasoro primary and its boundary-graviton tower
are therefore properties of the \emph{reduced} state space, not labels
that have to be supplied to the Wilson line from the outset.

\subsubsection{Discrete Series and its Affine Module}\label{sec:reps}

For an elliptic Wilson line the relevant horizontal representations are
the lowest-weight discrete series of $SL(2,\IR)$.  We use the
Cartan--Weyl basis $\{J^3,J^\pm\}$,
\beq\label{eq:sl2R-algebra}
  [J^3,J^\pm]=\pm J^\pm\,,
  \qquad
  [J^+,J^-]=-2J^3\, .
\eeq
The representation ${\cal D}^+_j$ is generated from an extremal vector
$|j,j\ket$ obeying
\beq\label{eq:D+j-conditions}
  J^3_0|j,j\ket=j|j,j\ket\,,
  \qquad
  J^-_0|j,j\ket=0\,,
\eeq
and has $J^3_0$ spectrum $j,j+1,j+2,\ldots$.  We work with the
universal cover, so $j$ need not be quantized.  With the Casimir
convention
\beq\label{eq:Casimir-discrete}
  \widehat C_2
  =-(J^3_0)^2
   +\frac12\bigl(J^+_0J^-_0+J^-_0J^+_0\bigr)\,,
  \qquad
  C_2({\cal D}^+_j)=-j(j-1)\, ,
\eeq
the Sugawara weight of every affine-primary state in this horizontal
multiplet is
\beq\label{eq:Sugawara-weight-discrete}
  \Delta_j=\frac{C_2({\cal D}^+_j)}{k-2}
  =-\frac{j(j-1)}{k-2}\, .
\eeq
The unreduced affine module is obtained by acting with the negative
current modes,
\beq\label{eq:D+j-basis}
  \widehat{\cal D}^{\,+}_j
  ={\rm span}\!\left\{
  (J^+_0)^p
  J^{a_1}_{-n_1}\cdots J^{a_q}_{-n_q}|j,j\ket
  \ ;\ p\geq0,\ n_i>0
  \right\} .
\eeq
The infinite horizontal tower is part of the unreduced Chern-Simons
state space.  

\subsubsection{The Half-Cylinder State}\label{sec:Hilbert-SL2}

At fixed Euclidean time the spatial slice is the punctured disc
$D^2\setminus\{0\}$, with the puncture labelled by $(j,\bar j)$.  Before
reduction the corresponding chiral state spaces are
\beq\label{eq:H-SL2}
  {\cal H}_{\rm aff}(D^2;j,\bar j)
  =\widehat{\cal D}^{\,+}_j
  \otimes
  \widehat{\bar{\cal D}}^{\,+}_{\bar j}\, .
\eeq
For a spinless defect we take $j=\bar j$.

The Euclidean path integral on
$M_-=D^2\times(-\infty,0]$, with a Wilson line running along the
Euclidean-time direction and ending on the puncture of the final
slice, prepares the corresponding Chern-Simons state.  At the
asymptotic cylinder we use the standard chiral Chern-Simons boundary
term and polarization for which the unreduced boundary degrees of
freedom carry the affine current algebra. Brown-Henneaux boundary
conditions are \emph{not} yet imposed.  Schematically,
\beq\label{eq:PI-state-SL2}
  |\Psi_{j,\bar j}\ket_{\rm aff}
  =
  \int_{\mathcal P_{\rm aff}}
  [\mathcal D A]\,[\mathcal D\widetilde A]\,
  W_{{\cal D}^{+}_j}[A]\,
  W_{\bar{\cal D}^{+}_{\bar j}}[\widetilde A]\,
  e^{-S_E[A,\widetilde A]}
  |A,\widetilde A\ket_{\tau=0}\, ,
\eeq
where $S_E$ denotes the Euclidean continuation of
\eqref{eq:CS-action}, including the boundary term.  The two Wilson factors refer to the independent
left and right Chern-Simons sectors. The bar on
$\bar{\cal D}^{+}_{\bar j}$ labels the right-moving copy, and no
complex conjugation of the gauge field is implied.

Throughout the barred copy we use the opposite Cartan--Weyl basis
$(\bar J^3,\bar J^+,\bar J^-)\equiv(-H,F,E)$.  Thus the superscript $+$ on
$\bar{\cal D}^{+}_{\bar j}$ denotes lowest weight with respect to this barred
basis, in particular $\bar J^-_0|\bar j,\bar j\ket=0$.  The barred basis obeys
the same abstract commutation relations as~\eqref{eq:sl2R-algebra}, with the
same invariant form $-2\Tr$, it also has $\bar\kappa^{33}=-1$ and
$\bar\kappa^{+-}=2$, so the affine conventions are identical in form.

Keeping the endpoint representation index open, the puncture supplies
the whole horizontal multiplet of affine-primary states.  For each
$v\in{\cal D}^+_j$ there is a level-zero state $|j;v\ket \in \widehat{\cal D}^{\,+}_j$ satisfying
\beq\label{eq:primary-condition-SL2}
  J^a_{n>0}|j;v\ket=0\,,
\eeq
and similarly in the barred sector.  Contracting the open endpoint
with a particular vector $v$ chooses one such upstairs state. Leaving
the index open keeps the construction representation-valued.  The
Brown-Henneaux reduction below removes this horizontal degeneracy and
selects the gravitational primary cohomologically.

\subsubsection{Semi-Classical Matching with the Deficit Angle}\label{sec:deficit-match}

The discrete-series label also fixes the classical elliptic holonomy.
To match the complex-diagonal form of the elliptic holonomy in
\eqref{eq:elliptic-holonomy}, and to make its relation to the
hyperbolic sector above threshold transparent, it is convenient to
place the corresponding factor of $i$ in the elliptic generator.  We
therefore define
\[
  H_{\rm ell}\equiv iH=\frac{i}{2}\sigma_3
\]
and parametrize the classical orbit charge as
$Q^{\rm orb}_{\rm ell}=2\rho H_{\rm ell}$.  The linking holonomy is then
\beq\label{eq:hol-j}
  {\rm Hol}(A)
  =\exp\!\left(\frac{2\pi}{k}Q^{\rm orb}_{\rm ell}\right)
  =
  \begin{pmatrix}
    e^{2\pi i \rho/k}&0\\
    0&e^{-2\pi i \rho/k}
  \end{pmatrix} .
\eeq
Matching this with \eqref{eq:elliptic-holonomy} gives the
classical relation $2\rho/k=\alpha$.  Upon quantization, the
coadjoint orbit becomes the representation carried by the Wilson
line, so its classical orbit parameter $\rho$ is correspondingly
mapped to the representation label $j$.  Standard coadjoint-orbit
quantization makes this relation precise:
$ \rho=j-\frac12$
\cite{AshokTroost}. Since the match to the geometry is by definition semiclassical, the relation we will use is
\beq\label{eq:j-alpha}
  j=\frac{k\alpha}{2}+O(1)\, .
\eeq
In particular, $j=k/2$ approaches the global AdS$_3$ endpoint $\alpha=1$ in the large-$k$ limit. The corresponding Brown-Henneaux weight
inferred from the classical geometry is
\beq\label{eq:h-check}
  h_{\rm grav}
  =\frac{c}{24}(1-\alpha^2)
  =\frac{k}{4}-\frac{j^2}{k}+O(1)\, ,
\eeq
where we treat $j \sim k$. We now derive the finite-$k$ counterpart of this discussion directly from the quantum constraint.

\subsection{Brown-Henneaux Reduction to Virasoro}
\label{sec:microstate-tower}

The reduction relevant for gravity is the one in the boundary
\emph{cylinder} frame.  This is important because the DS/Brown-Henneaux constraint on the cylinder and the frequently used
constant DS constraint on the plane correspond to different affine
modes.  They are related by one unit of spectral flow.

\subsubsection{The Cylinder Constraint and Spectral Flow}
\label{sec:DS-conical}

Brown-Henneaux boundary conditions are imposed directly on the boundary cylinder.  In the usual lowest-weight DS gauge they fix the coefficient of $F$ in
\beq\label{eq:DS-gauge-conical}
  a_\phi
  =F-\frac{2\pi}{k}\,\mathcal{L}(\phi)E
\eeq
to a nonzero constant, while the remaining field $\mathcal{L}(\phi)$ is the classical Virasoro stress tensor \cite{CHvD,CJ1}.  In the quantum current algebra, the classical $F$ direction becomes
the affine $J^-$ current\footnote{Note that the invariant form we use here has the opposite sign, compared to the one in Section~\ref{sec:CS-review}.}. Since $J^a_n$ are the Fourier modes of the boundary current, this immediately gives
\beq\label{eq:cylinder-vs-plane-constraint}
  J^-_{\rm cyl}(w)=1
  \quad\Longleftrightarrow\quad
  J^-_n=\delta_{n,0}\, .
\eeq
We have chosen the normalization to be unity. Its magnitude can be absorbed by a constant Cartan rescaling.  To compare with the standard plane discussion, let $w=\tau+i\phi$ and $z=e^w$.  Before the DS improvement the affine current has weight one, so
\beq\label{eq:plane-cylinder-current}
  J^-_{\rm pl}(z)=\sum_{n\in\IZ}J^-_n z^{-n-1}\,,
  \qquad
  J^-_{\rm cyl}(w)=zJ^-_{\rm pl}(z)
  =\sum_{n\in\IZ}J^-_n e^{-nw}\, .
\eeq
Thus the commonly used constant plane condition $J^-_{\rm pl}(z)=1$ fixes $J^-_{-1}=1$, while the Brown-Henneaux cylinder condition fixes $J^-_0=1$. 

In modes the first-class constraints are therefore
\beq\label{eq:DS-constraints-cylinder}
  \varphi_n=J^-_n-\delta_{n,0}\approx0\,,
  \qquad n\in\IZ\, .
\eeq
For the right-moving copy the identical-looking condition is
$\bar J^-_n=\delta_{n,0}$ in the barred basis above.  In the common matrix
basis this fixes the $E$ component of $\widetilde a_\phi$, since
$\bar J^-=E$, while the remaining component carrying
$\widetilde{\mathcal L}$ is $\bar J^+=F$.
Since $[J^-_m,J^-_n]=0$, their algebra is abelian without a quantum
central term.  The associated BRST charge is simply
\beq\label{eq:BRST-cylinder}
  Q_{\rm DS}
  =\sum_{n\in\IZ}c_{-n}
    \bigl(J^-_n-\delta_{n,0}\bigr)\,,
  \qquad Q_{\rm DS}^2=0\, .
\eeq

The cylinder and plane are related by one unit of spectral flow.  With
spectral-flow parameter $\omega$ the flow takes the standard form
\bea\label{eq:spectral-flow-conical}
  \widetilde J^3_n
  &=&J^3_n-\frac{k}{2}\omega\,\delta_{n,0}\,,
  \non\\
  \widetilde J^\pm_n
  &=&J^\pm_{n\pm\omega}\,,
  \non\\
  \widetilde L^{\rm Sug}_n
  &=&L^{\rm Sug}_n+\omega J^3_n
     -\frac{k}{4}\omega^2\delta_{n,0}\, .
\eea
It is immediate that for $\omega=-1$ one has $\widetilde J^-_{-1}=J^-_0$.  Thus the
Brown-Henneaux condition on the cylinder is a one-unit
spectrally flowed version of the plane DS constraint.  We will see that it fixes the physical Hamiltonian
below.  

This difference has direct representation-theoretic consequences.  At
level zero the cylinder reduction requires
$(J^-_0-1)|\psi\rangle=0$.
So a surviving state must have nonzero $J^-_0$ eigenvalue.  On any
finite-dimensional horizontal module, $J^-_0$ is nilpotent and therefore
cannot have eigenvalue one. Equivalently, $J^-_0-1$ is invertible, so the
level-zero BRST cohomology is empty.  This includes the trivial horizontal
module underlying the affine vacuum.  Because they are {\em infinite}-dimensional, the discrete
and principal-series modules relevant for us, evade this: in the
appropriate completion the same constraint can have nontrivial solutions,
which furnish the level-zero classes of the flowed reduction.

For standard quantum DS reduction and its BRST formulation see
\cite{FKW,dBT,BersOoguri,FeiginFrenkel,ArakawaVanishing,ArakawaReps}.
Spectrally flowed DS reduction is discussed by Dhillon
\cite{Dhillon}.  Because spectral flow changes the reduction functor, we
do not invoke the usual unflowed highest-weight vanishing theorems as
an all-level proof for the modules used here.  The flowed BRST complex,
including its extension to relaxed modules, is analyzed in detail in
\cite{Future}: below we retain only the ingredients needed in this
paper. 

\subsubsection{BRST Grading, Central Charge and Physical Weight}
\label{sec:BRST}

The DS improvement is
\beq\label{eq:DS-improved-stress}
  T_{\rm imp}=T_{\rm Sug}-\partial J^3\, ,
\eeq
under which $J^-$ has weight zero, as required for a nonzero constant
constraint.  Including the $(b,c)$ ghosts of weights $(0,1)$ gives
\beq\label{eq:DS-total-stress}
  T_{\rm tot}=T_{\rm Sug}-\partial J^3+T_{bc}\, .
\eeq
Its central charge is
\beq\label{eq:c-Liouville}
  c
  =\frac{3k}{k-2}+6k-2
  =1+6Q^2\,,
  \qquad
  Q=b+\frac1b\,,
  \qquad
  b^{-2}=k-2\, .
\eeq
This is the usual Liouville parametrization and tends to the
Brown-Henneaux result $c=6k$ at large $k$.

The BRST complex carries a useful level grading.  With $\Delta_j$ the
Sugawara weight we introduced earlier, define
\beq\label{eq:DS-level-conical}
  N=L_0^{\rm Sug}+L_0^{\rm gh}-\Delta_j\, .
\eeq
Since\footnote{The vanishing of this commutator can be seen directly mode by mode.
With respect to the Sugawara stress tensor, $J^-$ is a weight-one
current, and hence $[L_0^{\rm Sug},J^-_n]=-nJ^-_n$. Similarly, since $c$ has weight one with respect to the ghost stress
tensor, $[L_0^{\rm gh},c_n]=-n c_n$.
Therefore, using that the matter and ghost sectors commute,
$[L_0^{\rm Sug}+L_0^{\rm gh},\,c_{-n}J^-_n]
  =\bigl(n-n\bigr)c_{-n}J^-_n=0$.
The remaining subtraction term in $Q_{\rm DS}$ is $-c_0$, which also
commutes with $L_0^{\rm Sug}+L_0^{\rm gh}$.  Thus every term in
\eqref{eq:BRST-cylinder} has level zero, establishing the commutator
above.}
\[
  [Q_{\rm DS},L_0^{\rm Sug}+L_0^{\rm gh}]=0\,,
\]
one has $[Q_{\rm DS},N]=0$, and the cohomology can therefore be
computed level by level. 

A separate question is which Virasoro zero mode acts on this
cohomology.  An operator representing $L_0$ on the reduced theory must
preserve BRST-closed states (and BRST equivalence classes), so we seek
a representative commuting with $Q_{\rm DS}$.  The zero mode of the
spectrally unflowed DS stress tensor
$
  T_{\rm tot}$
is
$
  L_0^{\rm imp}+L_0^{\rm gh}
  =L_0^{\rm Sug}+J^3_0+L_0^{\rm gh}$.
For the cylinder BRST charge, however, this is not BRST invariant.
Indeed, using \eqref{eq:BRST-cylinder} and
$[J^-_n,J^3_0]=J^-_n$,
\beq\label{eq:BRST-L0-check}
  [Q_{\rm DS},J^3_0]=Q_{\rm DS}+c_0\, \
  \implies \
  [Q_{\rm DS},L_0^{\rm imp}+L_0^{\rm gh}]
  =Q_{\rm DS}+c_0\, \neq 0.
\eeq
Thus the unflowed improved zero mode does not in general preserve the
BRST-closed subspace.  This is precisely where the spectral flow of
the cylinder reduction matters: the stress tensor must be taken in
the same flowed frame as the current constraint.  With the ghost
sector unchanged, the corresponding improved total stress tensor is
\[
  \widetilde T_{\rm tot}
  =\widetilde T_{\rm Sug}-\partial\widetilde J^3+T_{bc}\, ,
\]
and we define its zero mode to be the physical Virasoro generator,
\bea\label{eq:L0-physical-conical}
  L_0^{\rm phys}
  &\equiv&
  \widetilde L_0^{\rm Sug}
     +\widetilde J^3_0+L_0^{\rm gh}
  \non\\
  &=&
  \left(L_0^{\rm Sug}-J^3_0-\frac{k}{4}\right)
     +\left(J^3_0+\frac{k}{2}\right)
     +L_0^{\rm gh}
  \non\\
  &=&
  L_0^{\rm Sug}+L_0^{\rm gh}+\frac{k}{4}\, .
\eea
In the last two lines we used the $\omega=-1$ spectral-flow
relations.  Since the final expression differs from
$L_0^{\rm Sug}+L_0^{\rm gh}$ only by a constant, it satisfies
\[
  [Q_{\rm DS},L_0^{\rm phys}]=0\,,
\]
and therefore defines the Virasoro zero mode on the BRST cohomology.

Note that the $J^3_0$ dependence cancels. We briefly explain why this is important. Before spectral flow, the improved zero mode $L_0^{\rm imp}$ contains an explicit
$J^3_0$ contribution, as we saw earlier. The level-zero BRST solution, however, must necessarily be a superposition of
states with different $J^3_0$ eigenvalues\footnote{Note that the BRST constraint $J^{-}_0|\psi\rangle = |\psi\rangle$ is incompatible with a definite $J_0^3$ eigenvalue.}. The rest of the terms in $L_0^{\rm imp}$ (the Sugawara term and the ghost piece) are fixed at level zero, within a fixed horizontal representation. So the only term whose eigenvalue varies across the different $J^3_0$ components is $J^3_0$ itself. This means that the BRST solution cannot be an eigenstate of the unflowed improved $L_0^{\rm imp}$. The flowed version on the other hand does not have $J^3_0$-dependence, and therefore this problem goes away. In other words: BRST
forces the state away from definite $J^3_0$ weights, but spectral flow comes to the rescue by removing $J^3_0$ from the physical $L_0$.

On level zero one obtains the following formula (which applies for general modules):
\beq\label{eq:h-Casimir-DS}
  h=\frac{C_2}{k-2}+\frac{k}{4}\, .
\eeq
For the discrete series this gives
\beq\label{eq:h-j-exact}
  h_j
  =\frac{k}{4}-\frac{j(j-1)}{k-2}
  =\frac{c-1}{24}
   -\frac{(j-\tfrac12)^2}{k-2}
  =\frac{Q^2}{4}+P_j^2\,,
  \qquad
  P_j=\frac{i(j-\tfrac12)}{\sqrt{k-2}}\, .
\eeq
Thus the elliptic branch has imaginary Liouville momentum and lies
below the threshold $(c-1)/24$.  On the branch\footnote{The reflected branch $j<\frac12$ is distinct at the unreduced
$SL(2,\IR)$ level, but corresponds to $\rho=j-\frac12<0$.  The associated
metric geometry and reduced Virasoro data are insensitive to
$\rho\to-\rho$. We therefore choose the branch $\rho>0$, equivalently
$j>\frac12$.} 
$j>\frac12$, the
interval\footnote{An interesting comparison is with the Maldacena--Ooguri
window $\frac12<j<\frac{k-1}{2}$ \cite{Maldacena:2000hw}.
There also, the no-ghost bound itself is $j<k/2$. The stronger upper
bound follows from requiring the worldsheet WZW spectrum to be
closed under spectral flow, since one unit of flow maps
$\widehat{\mathcal D}^{+}_{j}$ to
$\widehat{\mathcal D}^{-}_{k/2-j}$, together with the
$j>1/2$ lower bound from the zero-mode harmonic analysis.
That additional requirement is not imposed here: spectral flow here is part of the reduction defining
the gravitational state space.  Thus the Maldacena--Ooguri
tightening of the upper bound does not directly apply. The endpoint $j=k/2$ is instead singled out by
$h_j=0$, which eventually becomes the Virasoro vacuum.}
\beq\label{eq:j-bound}
  \frac12<j<\frac{k}{2}
\eeq
corresponds to $0<h_j<(c-1)/24$.  The upper endpoint is particularly
instructive:
\beq\label{eq:vacuum-endpoint-check}
  j=\frac{k}{2}
  \quad\Longrightarrow\quad
  h_j=0\,,
  \qquad
  P_j=\frac{i}{2}\left(b+\frac1b\right)=\frac{iQ}{2}\, .
\eeq
This is the $(1,1)$ degenerate Virasoro weight.  It is important, however, to distinguish the value of the highest weight from the Virasoro module itself.  The flowed DS reduction of the $j=k/2$ affine module gives us the full $h=0$ Verma module. It does not by
itself implement the level-one quotient that turns this into the Virasoro vacuum module.  But for the cylinder reduction done here, we believe it is quite natural to do this quotient.  

Firstly, it is natural that the $j=k/2$ sector (rather than say, the affine vacuum module ($j=0$)) is what corresponds to the Virasoro vacuum. This is what the bulk geometry requires: smooth global AdS$_3$ has central
spatial holonomy, invisible in the metric -- in the fundamental
representation \eqref{eq:elliptic-holonomy} it is $-\mathbf 1$ -- and
in the semiclassical dictionary of Section~\ref{sec:deficit-match} it
is the $j=k/2$ sector that carries it.\footnote{Combining the exact
relation $\rho=j-\frac12$ with the classical identification
$\alpha=2\rho/k$ gives $\alpha=1-1/k$ at $j=k/2$ rather than
$\alpha=1$ on the nose.  This residual $O(1/k)$ is below the
resolution of the semiclassical matching \eqref{eq:j-alpha}.  The
exact finite-$k$ statement is that the flowed DS reduction reaches
$h=0$ (equivalently, $P=iQ/2$) at $j=k/2$. The vacuum module requires
the additional level-one quotient.} This is consistent also with the fact that the vacuum has extra isometries which need quotienting, as reflected in the orbit language: it corresponds to the exceptional orbit.  Thus the absence of the Wilson line should not be identified with the
gravitational vacuum in this reduction, and there is no contradiction in the affine vacuum module having trivial BRST cohomology.

There is also a natural ``upstairs'' motivation for the additional
quotient.  Precisely at $j=k/2$ the affine module develops a singular
sub-module.  The flowed BRST reduction is nevertheless blind to
this reducibility: it still produces the full
$h=0$ Virasoro Verma module \cite{Future}.  But if one views the quotiented module upstairs as the more fundamental object,  it is natural to quotient by the singular sub-module downstairs as well. And there is indeed such a candidate at $h=0$:  the level-one singular sub-module generated by $L_{-1}|0\rangle$.  The
result is the Virasoro vacuum module.  Thus the vacuum is obtained by
first reaching the $h=0$ endpoint of the generic family and only then
performing the extra quotient.

This story-arc gives further support for the ``regulate-and-quotient''
prescription of \cite{CKPSP}, in which the vacuum is treated as a
limiting normal orbit and the quotient associated with its enhanced
stabilizer is imposed only afterwards.  In particular, the normal-orbit
quantization gives a single central charge
$c=C+1$ across the sectors, where $C$ is the classical central charge
of the orbit.  By \eqref{eq:vacuum-endpoint-check},
\[
  P^2\big|_{j=k/2}
  =-\frac{Q^2}{4}
  =-\frac{c-1}{24}\, ,
\]
while the exceptional vacuum orbit has
$2\pi b_0=-C/24$.  Using the normal-orbit relation
$2\pi b_0=h-(c-1)/24$, the two agree precisely when $C=c-1$.
This should be contrasted with the $c=C+13$ result obtained by
quantizing the exceptional orbit as an independent object
\cite{CJ1, GMY}.  The correspondence is therefore quite direct:
generic $j$ in the interval \eqref{eq:j-bound} reduces to a Virasoro
Verma module, $j=k/2$ reaches its $h=0$ endpoint, and the subsequent
level-one quotient produces the vacuum module.  The singular
sub-module that appears simultaneously in the $j=k/2$ affine module
provides an independent indication that precisely this endpoint
requires special quotienting \cite{Future}.

This also gives a sharp check on the cylinder reduction: applying the
unflowed plane improvement to the original lowest-weight vector would
instead give
\beq\label{eq:plane-weight-check}
  h_{\rm pl}=\Delta_j+j
  =\frac{j(k-1-j)}{k-2}\,,
  \qquad
  h_{\rm pl}\big|_{j=k/2}=\frac{k}{4}\,,
\eeq
which does not reproduce the vacuum endpoint.  Finally,
inserting the semiclassical identification \eqref{eq:j-alpha} into
\eqref{eq:h-j-exact} gives
\beq\label{eq:h-j-semiclassical}
  h_j
  =\frac{k}{4}-\frac{j^2}{k}+O(1)
  =\frac{c}{24}(1-\alpha^2)+O(1)\, ,
\eeq
in agreement with \eqref{eq:h-check}.  The gravitational weight is
therefore an output of the flowed Brown-Henneaux reduction.

\subsubsection{A Single Primary and the Virasoro Tower}
\label{sec:zero-mode-conical}

It remains to see what happens to the infinite horizontal multiplet.
At level zero and ghost number zero, the BRST condition
\eqref{eq:BRST-cylinder} is simply
\beq\label{eq:zero-mode-constraint-conical}
  (J^-_0-1)|\psi\ket=0\, .
\eeq
For the discrete series this equation has an honest solution in the
Hilbert-space completion of the horizontal module.  Writing
\beq\label{eq:zero-mode-representative-conical}
  |\psi\ket
  =\sum_{p\geq0}c^W_p(J^+_0)^p|j,j\ket\, ,
\eeq
and using
\beq\label{eq:Whittaker-recursion-conical}
  J^-_0(J^+_0)^p|j,j\ket
  =p(2j+p-1)(J^+_0)^{p-1}|j,j\ket\, ,
\eeq
one finds
\beq\label{eq:zero-mode-recursion-conical}
  (p+1)(2j+p)c^W_{p+1}=c^W_p\,,
  \qquad
  c^W_p=\frac{c^W_0}{p!\,(2j)_p}\, .
\eeq
For $j>0$ the denominator never vanishes, so the solution is unique up
to its overall normalization.  Moreover, with the standard unitary
discrete-series norm,
$\|(J^+_0)^p|j,j\ket\|^2=p!\,(2j)_p$, and hence
\beq\label{eq:zero-mode-norm-conical}
  \langle\psi|\psi\rangle
  =|c^W_0|^2\sum_{p=0}^\infty
    \frac{1}{p!\,(2j)_p}
  =|c^W_0|^2\,{}_0F_1(;2j;1)<\infty\, .
\eeq
Thus the nonzero cylinder constraint selects a single level-zero
state. In the discrete series, it is in addition,
normalizable.  We use the ghost vacuum convention
$b_{n\geq0}|0\ket_{\rm gh}=0$ and
$c_{n\geq1}|0\ket_{\rm gh}=0$, so the level-zero ghost sector is
spanned by $|0\ket_{\rm gh}$ and $c_0|0\ket_{\rm gh}$ and contains
no ghost-number $-1$ state.  This closed state therefore supplies the
unique ghost-number-zero BRST class.  It is the Virasoro primary of
weight \eqref{eq:h-j-exact}.  The principal-series analogue in the next section is instead represented
by a formal, non-normalizable Whittaker vector in the appropriate
completion. The barred zero-mode argument is identical in the barred basis, with
$J^\pm$ replaced by $\bar J^\pm$.

At positive level the reduction is explicitly cohomological: the
constraint and its gauge symmetry are expected to reorganize the three
affine-current towers into the single Virasoro tower generated by the
reduced stress tensor.  For the \emph{flowed} reduction used here, the
all-level statement is not being inferred directly from the standard
unflowed highest-weight vanishing theorems.  The detailed BRST analysis
will be presented in \cite{Future}.  The all-level statement used in the
remainder of this paper for generic $j$ is
\beq\label{eq:DS-map-conical}
  H^0_{\rm DS,flow}\!\left(\widehat{\cal D}^{\,+}_j\right)
  \cong {\cal V}_{h_j}\, .
\eeq
What has been established directly above is the level-zero part of
this statement and the physical grading and weight. We have checked the cohomology explicitly at low levels and various ghost numbers, but a full discussion (for both discrete and continuous series) is part of \cite{Future}.  With that
reduction understood, the physical descendants are
\beq\label{eq:microstate-tower}
  L_{-n_1}\cdots L_{-n_p}\,
  \bar L_{-m_1}\cdots\bar L_{-m_q}\,
  |h_j,\bar h_j\ket\,,
  \qquad n_i,m_i>0\, .
\eeq
Their levels are measured by the BRST-invariant grading
\eqref{eq:DS-level-conical}.

\subsection{Summary and a Look Ahead}\label{sec:summary-look-ahead}

Let us summarize what we have done. An elliptic Wilson line prepares the unreduced puncture sector
$\widehat{\cal D}^{\,+}_j\otimes
\widehat{\bar{\cal D}}^{\,+}_{\bar j}$ on the spatial disc.  Its
holonomy fixes the semiclassical defect parameter through
$j=k\alpha/2+O(1)$.  Brown-Henneaux boundary conditions then impose a
nonzero constant current on the boundary cylinder, which we
normalize to $J^-_0=1$.  Relative to the common plane convention this is a
one-unit spectrally flowed DS reduction.  The flow fixes the physical
Hamiltonian and gives
\beq\label{eq:summary-flowed-weight}
  L_0^{\rm phys}
  =L_0^{\rm Sug}+L_0^{\rm gh}+\frac{k}{4}\,,
  \qquad
  h=\frac{C_2}{k-2}+\frac{k}{4}\, .
\eeq
For $C_2=-j(j-1)$ this is the exact sub-threshold weight
\eqref{eq:h-j-exact}: the level-zero constraint has a unique
normalizable solution in the completed discrete-series module, and
$j=k/2$ leads to the Virasoro vacuum.  

The same Brown-Henneaux reduction will be used in the hyperbolic
sector.  What changes there is the horizontal representation: the
principal continuous series is relaxed and has no extremal vector.
The module-independent part of the calculation above nevertheless
continues immediately.  In particular, for
$C_2=s^2+\frac14$ one obtains
\beq\label{eq:summary-continuous-weight}
  h_P=\frac{c-1}{24}+P^2\,,
  \qquad
  P=\frac{s}{\sqrt{k-2}}\, .
\eeq
The existence and uniqueness of the level-zero class can again be
shown directly, while the all-level cohomology of the relaxed module
requires additional care.  We will use the result in the next section
and defer the detailed BRST analysis to \cite{Future}.

\section{BTZ Microstates and Super-Threshold Primaries}\label{sec:super-threshold}

We now turn to the main construction of this paper: the bulk state
associated with a Virasoro primary above the BTZ threshold.  The logic is
the same as in the elliptic sector of Section~\ref{sec:microstates}, with
one important change upstairs.  The Wilson line now carries a principal
continuous-series representation, appropriate to hyperbolic holonomy, so
the unreduced puncture sector is a relaxed affine module.  Brown-Henneaux
boundary conditions then impose the same cylinder reduction derived
in Section~\ref{sec:microstate-tower}: the nonzero constant constraint is
$J^-_0=1$, and relative to the plane constraint this is a
one-unit spectrally flowed Drinfel'd-Sokolov reduction.

Thus the quantum construction has the schematic form
\beq\label{eq:super-threshold-logic}
  \hbox{principal-series Wilson line}
  \longrightarrow
  \widehat{\mathcal C}^{\,0}_s
  \xrightarrow{\;H^0_{\rm DS,flow}\;}
  \mathcal V_{h_P}\, .
\eeq
The Wilson line fixes the hyperbolic conjugacy class of the spatial
holonomy.  The flowed reduction removes the unreduced horizontal
zero-mode label and produces the Virasoro primary together with its
boundary-graviton descendants.  The module-independent part of this
reduction, including the physical Hamiltonian and central charge, was
already derived in Section~\ref{sec:microstate-tower}.  Here we only need
to explain how it acts on the principal continuous series.

The resulting primary has real Liouville momentum $P$ and weight
\beq
  h_P=\frac{c-1}{24}+P^2\, ,
\eeq
and therefore lies above the threshold $(c-1)/24$.  Distinct heavy
primaries, rather than their universal Virasoro descendants, are the
microscopic labels relevant to the black-hole entropy.  The descendants
are subleading boundary-graviton excitations built on each such primary \cite{CKPSP}.

\subsection{BTZ Geometry and Hyperbolic Holonomy}\label{sec:BTZ-holonomy}

For illustration, consider first the non-rotating BTZ geometry \cite{BTZ} in
Euclidean signature,
\beq\label{eq:BTZ-metric}
  ds^2
  =
  (r^2-r_+^2)\,d\tau^2
  +
  \frac{dr^2}{r^2-r_+^2}
  +
  r^2\,d\phi^2,
  \qquad
  \phi\sim\phi+2\pi\, .
\eeq
When $\tau$ is periodically identified with inverse temperature
$\beta=2\pi/r_+$, this is the smooth Euclidean BTZ saddle.  In the
present section we use the metric only to read off the classical
holonomy associated with a heavy primary sector \cite{CKPSP}.  Without
the thermal identification, the geometry will be singular at the horizon. We will discuss this in detail, later.

The ADM mass is
\beq
  M=\frac{r_+^2}{8G_N}\, ,
\eeq
and the classical Brown-Henneaux weights are
\beq\label{eq:h-BTZ}
  h_L=h_R=\frac{c}{24}\left(1+r_+^2\right)\, .
\eeq
At finite $k$ the exact reduced Virasoro weight will instead be written
as $h_P=(c-1)/24+P^2$.  The difference between $c/24$ and
$(c-1)/24$ is an order-one quantum shift and is subleading in the
semiclassical large-$c$ regime.

In Chern-Simons language the gauge-invariant datum carried by the
Wilson line is the conjugacy class of the holonomy around a small
spatial circle linking it.  In the BTZ sector this holonomy is
hyperbolic. For one chiral copy we write
\beq\label{eq:hyperbolic-holonomy}
  {\rm Hol}(A)
  \sim
  \begin{pmatrix}
    e^{\pi\lambda} & 0 \\
    0 & e^{-\pi\lambda}
  \end{pmatrix},
  \qquad
  \lambda>0\, .
\eeq
For the non-rotating geometry $\lambda=r_+$.  The corresponding
Wilson-line orbit is therefore hyperbolic. A smooth black hole would correspond to the demand that the thermal holonomy is trivial, in the Chern-Simons language.

\subsection{Principal Continuous Series and the Flowed Reduction}\label{sec:continuous-reps}

The horizontal representation appropriate to a hyperbolic orbit is the
principal continuous series of $SL(2,\IR)$.  We denote it by
$\mathcal C_s^\epsilon$, with $s>0$ and
$J^3_0$ spectrum $m\in\epsilon+\IZ$.  The global label $\epsilon$ will
play no role below\footnote{Note however that it {\em is} part of the Wilson line label. It however does not enter the holonomy conjugacy class, and the reduction below is
carried out at fixed $\epsilon$ and returns the same Virasoro momentum
$P$ and an isomorphic Virasoro module for every value of $\epsilon$. We will view this as the statement that AdS$_3$ gravity is sensitive only to the reduced labels: it is a Virasoro theory, not an $SL(2,\IR)$ theory. But there may be some physics here that is of interest, and worth coming back to.}, and we fix $\epsilon=0$.  With the Casimir
convention of Section~\ref{sec:reps},
\beq\label{eq:Casimir-C}
  C_2(\mathcal C_s^\epsilon)
  =s^2+\frac14\, .
\eeq
Equivalently one may write $j=\frac12+is$, but the invariant label $s$
will be more useful.

There is no highest- or lowest-weight vector in $\mathcal C_s^0$.
The unreduced affine module is therefore the relaxed module induced
from the full horizontal representation,
\beq\label{eq:C-affine-basis}
  \widehat{\mathcal C}^{\,0}_s
  =
  {\rm span}\left\{
  J^{a_1}_{-n_1}\cdots J^{a_p}_{-n_p}\,v
  \; ;\;
  n_i>0,\; v\in\mathcal C^0_s
  \right\}.
\eeq
This is the natural sector produced by the Wilson line before
Brown-Henneaux reduction.  In particular, its infinite horizontal
multiplet is unreduced current-algebra data.

The gravitational reduction is exactly the cylinder reduction already
derived in Section~\ref{sec:DS-conical}.  Classically the corresponding
Drinfel'd-Sokolov form of the boundary connection is
\beq\label{eq:DS-gauge-main}
  a_\phi
  =
  F
  -
  \frac{2\pi}{k}\,\mathcal{L}(\phi)\,E\, ,
\eeq
while quantum mechanically the constraints are
$J^-_n=\delta_{n,0}$.  Because the current has weight one before the DS
improvement, a constant on the boundary cylinder fixes $J^-_0$, whereas
a constant in the standard plane presentation fixes $J^-_{-1}$.  As
shown in Section~\ref{sec:DS-conical}, the two are related by spectral
flow with $\omega=-1$. The corresponding
flowed BRST complex is developed in \cite{Future}. A noteworthy fact is that the level-by-level cohomology calculations in discrete and continuous series are largely parallel (summarized at the end of Appendix~\ref{app:DS-reduction}, details in \cite{Future}), and differ only in the value of the Casimir. 

For the present paper, the crucial module-independent result was already
obtained in eq.~\eqref{eq:h-Casimir-DS}:
\beq\label{eq:principal-flowed-weight}
  h
  =
  \frac{C_2}{k-2}+\frac{k}{4}\, .
\eeq
Substituting \eqref{eq:Casimir-C} gives
\beq\label{eq:Liouville-P}
  P=\frac{s}{\sqrt{k-2}}\, ,
\eeq
and hence\footnote{These expressions have also been presented in \cite{Blommaert}.}
\beq\label{eq:h-P}
  h_P
  =
  \frac{k}{4}+\frac{s^2+\tfrac14}{k-2}
  =
  \frac{Q^2}{4}+P^2
  =
  \frac{c-1}{24}+P^2,
  \qquad
  Q=b+\frac1b,
  \qquad
  b^{-2}=k-2\, .
\eeq
Thus the super-threshold weight is a direct output of the flowed
Brown-Henneaux reduction.  No $J^3_0$ eigenvalue appears: the
$J^3_0$ dependence cancels from the physical Hamiltonian under the
spectral flow, exactly as in Section~\ref{sec:microstate-tower}.

What remains module-dependent is the emergence of the primary class and
the structure of its positive-level cohomology.  At level zero the
answer is elementary.  In the natural completion of the principal
series, the nonzero-eigenvalue condition
\beq\label{eq:principal-zero-mode-main}
  (J^-_0-1)|\psi\rangle=0
\eeq
has a one-dimensional space of solutions for fixed $s$: $J^-_0$ shifts
the two-sided $J^3_0$ lattice with a coefficient that never vanishes for
$s>0$.  Appendix~\ref{app:DS-reduction} presents the short discussion.  Thus
no individual horizontal label $m$ survives as a label of the reduced
primary.  At positive affine level the statement is 
cohomological.  The all-level statement we expect is
\beq\label{eq:DS-map-main}
  H^0_{\rm DS,flow}\!\left(\widehat{\mathcal C}^{\,0}_s\right)
  \cong
  \mathcal V_{h_P}
\eeq
for generic real $P$.  As in the discrete sector, this is a
reorganization of the affine module by cohomology \cite{Future}.

\subsection{Matching to BTZ Parameters}\label{sec:BTZ-match}

For the hyperbolic orbit, take
$H_{\rm hyp}=\frac12\sigma_3$ and
$Q^{\rm orb}_{\rm hyp}=2\rho_{\rm hyp}H_{\rm hyp}$.  Then
eq.~\eqref{eq:flat-Aphi} gives holonomy eigenvalues
$e^{\pm2\pi\rho_{\rm hyp}/k}$.  Standard coadjoint-orbit
quantization gives the exact continuous-series dictionary
$\rho_{\rm hyp}=s$ \cite{AshokTroost}.  This fits naturally with the
elliptic result $\rho=j-\frac12$.  At the level of the complexified
orbit radius, the continuation
$\rho^2\to-\rho_{\rm hyp}^2$ (equivalently
$\rho\to\pm i\rho_{\rm hyp}$) takes the elliptic holonomy conjugacy
class into the hyperbolic one, and correspondingly
\[
  C_2=\frac14-\rho^2
  \ \longrightarrow\
  \frac14+\rho_{\rm hyp}^2
  =\frac14+s^2,
\]
in agreement with the principal-series Casimir
\eqref{eq:Casimir-C}.  Comparing the hyperbolic orbit holonomy with
the classical BTZ holonomy \eqref{eq:hyperbolic-holonomy} then gives
$\rho_{\rm hyp}=k\lambda/2$ semiclassically. This means that
\beq\label{eq:s-lambda}
  s=\frac{k\lambda}{2}+O(1),
  \qquad
  \lambda=\frac{2s}{k}+O(1/k)\, .
\eeq
For the non-rotating BTZ geometry,
$\lambda=r_+$, and therefore
\beq\label{eq:M-s}
  M
  =
  \frac{r_+^2}{8G_N}
  =
  \frac{s^2}{2k^2G_N}\left(1+O(1/k)\right)\, .
\eeq
Using $c=6k$ at leading semiclassical order, the Brown-Henneaux weight
is
\beq\label{eq:h-s}
  h
  =
  \frac{c}{24}\left(1+\lambda^2\right)
  =
  \frac{c}{24}+\frac{s^2}{k}+O(1)\, .
\eeq
This is precisely the large-$k$ limit of the exact reduced expression
\eqref{eq:h-P}.  

For a rotating BTZ sector the two chiral Wilson lines carry independent
labels $s_L,s_R$, or equivalently $P_L,P_R$, with
\beq\label{eq:BTZ-L0-main}
  L_0-\frac{c}{24}
  =
  \frac{\ell M+J}{2},
  \qquad
  \bar L_0-\frac{c}{24}
  =
  \frac{\ell M-J}{2}\, ,
\eeq
and semiclassically
\beq\label{eq:rotating-match}
  s_L=\frac{k\lambda_L}{2}+O(1),
  \qquad
  s_R=\frac{k\lambda_R}{2}+O(1),
  \qquad
  \lambda_L=\frac{r_+ + r_-}{\ell},
  \qquad
  \lambda_R=\frac{r_+ - r_-}{\ell}\, .
\eeq
The exact reduced weights are
$h_L=(c-1)/24+P_L^2$ and
$h_R=(c-1)/24+P_R^2$, with
$P_{L,R}=s_{L,R}/\sqrt{k-2}$.

\subsection{The Path-Integral State}\label{sec:Hilbert-super}

Before reduction, the punctured-disc state space in the two chiral
sectors is
\beq\label{eq:H-super}
  \mathcal H_{\rm aff}(D^2;s_L,s_R)
  =
  \widehat{\mathcal C}^{\,0}_{s_L}
  \otimes
  \widehat{\bar{\mathcal C}}^{\,0}_{s_R}\, .
\eeq
The barred principal-series module uses the same opposite-Borel
convention fixed in Section~\ref{sec:Hilbert-SL2}.
The Euclidean path integral on the half-cylinder
$M_-=D^2\times(-\infty,0]$, with principal-series Wilson lines ending on
the puncture of the final slice, prepares an unreduced affine state
schematically of the form
\beq\label{eq:PI-state-super}
  |\boldsymbol{\Psi}_{s_L,s_R}\rangle_{\rm aff}
  =
  \int_{\mathcal P_{\rm aff}}
  [\mathcal D A]\,[\mathcal D\widetilde A]\;
  W_{\mathcal C^0_{s_L}}[A]\,
  W_{\bar{\mathcal C}^0_{s_R}}[\widetilde A]\,
  e^{-S_E[A,\widetilde A]}
  |A,\widetilde A\rangle_{\tau=0}\, .
\eeq
As in Section~\ref{sec:Hilbert-SL2}, the open Wilson-line endpoint is
representation-valued.  Here there is no distinguished extremal vector
in the horizontal principal series, so the half-cylinder does not
canonically select a particular $J^3_0$ basis vector.  This causes no
ambiguity in the Brown-Henneaux theory: the entire affine sector is the
input to the flowed DS reduction, and the one-dimensional level-zero
cohomology supplies the Virasoro primary.

After imposing the Brown-Henneaux constraint in both chiralities, the reduction acts on the affine puncture sector,
\beq\label{eq:H-phys-super}
  H^0_{\rm DS,flow}\!\left(\mathcal H_{\rm aff}(D^2;s_L,s_R)\right)
  \cong
  \mathcal V_{h_L}\otimes\bar{\mathcal V}_{h_R}
  =
  \mathcal H_{\rm phys}(D^2;P_L,P_R)\, .
\eeq
The Wilson-line half-cylinder supplies the unreduced puncture sector entering this BRST complex, while imposing the Brown-Henneaux reduction prepares the physical cohomology class.  At level zero this is the primary $|P_L,P_R\rangle_{\rm phys}=|h_L\rangle_L\otimes|h_R\rangle_R$.  For $P_L=P_R=P$ it is the spinless heavy primary $|h_P,h_P\rangle$.  This is the precise sense in which the principal-series Wilson line constructs a super-threshold Virasoro primary: the Wilson line fixes the unreduced orbit and holonomy, while the physical primary is selected by the flowed Brown-Henneaux reduction.

\subsection{Virasoro Descendants}\label{sec:microstate-tower-super}

The reduced primary carries the usual Virasoro descendant tower,
\beq\label{eq:microstate-tower-super}
  \mathcal V_{h_L}\otimes \bar{\mathcal V}_{h_R}
  =
  {\rm span}\left\{
  L_{-n_1}\cdots L_{-n_p}\,
  \bar L_{-\bar n_1}\cdots \bar L_{-\bar n_q}
  |h_L,h_R\rangle
  \; ;\;
  n_i,\bar n_i>0
  \right\}.
\eeq
These are boundary-graviton excitations of a fixed primary sector.  A
descendant at levels $(N,\bar N)$ has
$L_0=h_L+N$ and $\bar L_0=h_R+\bar N$, so the Wilson line fixes the
primary holonomy data rather than (the exact energy of) every descendant. The latter are controlled by the non-zero-mode Virasoro generators, which act only on the boundary: they are boundary hair. The BRST-invariant level grading is the same one introduced in
\eqref{eq:DS-level-conical}.

The super-threshold construction may therefore be summarized as
\beq\label{eq:summary-table}
  \boxed{
  \mathcal C_s^0
  \ \longrightarrow\ 
  \widehat{\mathcal C}^{\,0}_s
  \ \xrightarrow{\;H^0_{\rm DS,flow}\;}\ 
  \mathcal V_{h_P},
  \qquad
  h_P=\frac{c-1}{24}+P^2,
  \qquad
  P=\frac{s}{\sqrt{k-2}}\, .}
\eeq
The subsequent sections study the semiclassical saddle associated with
this individual heavy primary and then its relation to the smooth
thermal black hole.

\section{The Semi-Classical Microstate: A Singular
  Horizon}\label{sec:semiclassical}

In Sections~\ref{sec:microstates} and~\ref{sec:super-threshold}, we
constructed the quantum state $|P\ket$ as a non-perturbative object
in Chern-Simons gravity by starting from the Euclidean half-cylinder path
integral with a continuous-series Wilson line at the origin of the disc and
imposing the Brown-Henneaux reduction.  In this section, we study
what this state looks like in the  large-$k$
limit, which we will view as a type of (semi-)classicality\footnote{But note the final bullet point in the Introduction.}.
Throughout this section, relations between the exact
representation label and geometric quantities are understood at leading
large $k$, and the subleading corrections in \eqref{eq:s-lambda} are
suppressed.

The result is as follows.  In the large-$k$ WKB limit, the
path integral~\eqref{eq:PI-state-super} is dominated by a saddle
point.  The saddle-point connection is determined by three
ingredients -- chirality demand on the gauge field, the holonomy fixed by the representation label $s$, and
the Drinfel'd-Sokolov\footnote{We will view chirality together with Drinfel'd-Sokolov gauge, as the definition of Brown-Henneaux boundary conditions. Note that our construction canonicalizes this notion of ``asymptotically AdS$_3$".} boundary conditions.  Translating from the gauge-field language to
the metric language via the dictionary~\eqref{eq:CS-connections}, the
saddle is the BTZ metric~\eqref{eq:BTZ-metric} with horizon radius
$r_+ = \lambda = 2s/k$.  However the microstate saddle has a fundamentally
different global structure from the thermal Euclidean BTZ\footnote{The discussions in this section straightforwardly generalize to rotating BTZ as well, but we restrict to the non-rotating case for clarity.}.  The
Euclidean time direction is \emph{non-compact}, there is no
periodicity condition, and the geometry terminates at the horizon with
a Wilson line source.  In the metric language, the horizon has a finite perimeter $2\pi r_+$,
but it is naturally viewed as a singular (geodesically incomplete) geometry with a source. If one lets this microstate run in a thermal loop for some fixed external temperature $\beta$, there is a codimension-2 conical excess/defect singularity at the horizon location, with smoothness only at the fine-tuned value $\beta=2\pi/r_+$. We emphasize that these are all semi-classical statements at large-$k$ and the state is well-defined as a quantum state at finite-$k$ for any $\beta$.

In Lorentzian signature, this singularity persists: the
semi-classical geometry of the individual microstate is the
exterior BTZ metric with a singular horizon that has no natural
extension into a smooth Kruskal diagram.  The smooth horizon is
recovered only as the property of an ensemble of states, as we will see in this and the next section.

These observations were anticipated in~\cite{CKPSP}
(see especially Section~4.2 of that reference), where the relevant Euclidean geometry was described in the Chern-Simons framework.  The present section provides the explicit saddle-point statement in the context of the path-integral construction of the microstate. The solutions are simple and live in the vicinity of well-known expressions. But we will be deliberately slow, because the boundary/regularity conditions are non-standard and we wish to be fully aware of assumptions, when we are making them.

\subsection{The WKB Limit}\label{sec:WKB}

The microstate is defined by the bulk Chern-Simons path
integral~\eqref{eq:PI-state-super} on the half-cylinder
$M_-=D^2\times(-\infty,0]$, with a continuous-series Wilson line along
the time axis at the centre of the disc.  The integration variables
are the bulk gauge fields $A$ and $\tilde A$, and it is this object that we
analyse at large $k$.

Since the action~\eqref{eq:CS-action} carries an overall factor of the
level $k$, the integral is controlled by a stationary-phase
approximation\footnote{One might worry about the uniqueness of the saddle. The only gauge-invariant data (or allowed gauge-variant data, which can only happen at boundary) are the holonomy controlled by the source (and the boundary gravitons). This is a standard fact about semi-classical bulk, and our statements will be precisely compatible with that. We will ignore boundary gravitons, they can be added without any change in our horizon/singularity discussions.}.  With $(A_*,\tilde A_*)$ the dominant stationary
configuration,
\beq\label{eq:WKB}
  |\boldsymbol{\Psi}_{s_L,s_R}\rangle_{\rm aff}
  \;\sim\;
  e^{-S_E[A_*,\tilde A_*]}\;
  \bigl|A_*,\tilde A_*\bigr\rangle_{\tau=0}\;
  \Bigl(1+\mathcal{O}(1/k)\Bigr)\, ,
\eeq
and the classical geometry of the microstate is the one built from
$(A_*,\tilde A_*)$.  The saddle solves the Chern-Simons equations of
motion on $M_-$ in the presence of the line source, subject to:
\begin{itemize}
\item \textbf{Wilson line source and holonomy:}
  by~\eqref{eq:CS-eom-source} the field strength of each connection
  has a delta function on the line and vanishes away from it.  The
  connection is flat on the punctured disc, and the conjugacy class of
  the holonomy linking the puncture is fixed by the Wilson-line label:
  for the principal continuous series it is hyperbolic with boost
  parameter $\lambda=2s/k$, by~\eqref{eq:s-lambda}.
\item \textbf{Boundary conditions:} the Brown-Henneaux conditions are
  imposed at the asymptotic boundary.  We take these to be chirality
  of the two connections together with the Drinfel'd-Sokolov
  gauge~\eqref{eq:DS-gauge-main}.
\item \textbf{Global structure:} the Euclidean time direction runs
  over the non-compact range $(-\infty,0]$.  No periodic
  identification is imposed, because none is part of the definition
  of~\eqref{eq:PI-state-super}.
\end{itemize}
Two comments.  First, the path integral~\eqref{eq:PI-state-super} does
not select a particular vector in the principal-series zero-mode
multiplet, but the saddle sees the Wilson line only through its
coadjoint orbit, and hence only through the Casimir and the holonomy
class.  Second, the delta function above is one in the Chern-Simons
field strength, and we will often refer to it as the
delta-function singularity in the curvature.  Its metric translation
depends on the global choices we make, and will also be discussed. Note that the singularity in the metric language is also technically a curvature singularity. But it is not a curvature divergence singularity as in Schwarzschild, it is a delta function defect type singularity.

\subsection{The Saddle Connection}\label{sec:saddle-connection}

We work in Lorentzian signature in this subsection and continue to
Euclidean signature at the end using $t=-i\tau$.

\medskip\noindent\textbf{Fefferman--Graham gauge.}\quad
We adopt the radial gauge 
\beq\label{eq:FG-gauge}
  A_\rho \;=\; H\,,
  \qquad
  \tilde{A}_\rho \;=\; -H\,,
\eeq
where $\rho$ is the radial coordinate on the disc, increasing from the
puncture out to the asymptotic boundary at $\rho\to\infty$.  Flatness
in the radial direction, $\de_\rho A_i+[H,A_i]=0$ for $i=\phi,t$,
fixes the $\rho$-dependence of the remaining components in terms of a
$\rho$-independent reduced connection $a_i$,
\beq\label{eq:FG-rho-dep}
  A_i(\rho) \;=\; e^{-\rho\,\text{ad}_{H}}\,a_i
  \;=\; (a_i)_E\,e^{-\rho}\,E
  \;+\; (a_i)_H\,H
  \;+\; (a_i)_F\,e^{+\rho}\,F\, ,
\eeq
where we used $[H,E]=E$ and $[H,F]=-F$, and similarly for
$\tilde{A}_i$ with $\rho\to-\rho$.  The Drinfel'd-Sokolov
form~\eqref{eq:DS-gauge-main} of the Brown-Henneaux conditions removes
the $H$ component and normalises the leading term of $a_\phi$ to
unity in the $F$ direction \cite{CJ1},
\beq\label{eq:reduced-a}
  a_\phi \;=\; -\frac{2 \pi}{k}\mathcal{L}\,E \;+\; F\,,
\eeq
with the constant $\mathcal{L}$, free.  For the
second connection the same conditions apply with $E$ and $F$
interchanged, reflecting the opposite chirality, so that
$\tilde a_\phi=-\frac{2\pi}{k}\widetilde{\mathcal{L}}\,F+E$.
This is precisely the barred convention of
Section~\ref{sec:Hilbert-SL2}: $\bar J^-=E$ is the fixed DS component, while
$\bar J^+=F$ carries $\widetilde{\mathcal L}$.

\medskip\noindent\textbf{Holonomy constraint.}\quad
The holonomy around the $\phi$-circle at any fixed $\rho$ is conjugate
to $\exp(2\pi a_\phi)$, independently of $\rho$.  In the fundamental
representation, using the choices fixed in
Section~\ref{sec:CS-review}, the
matrix $a_\phi$ in~\eqref{eq:reduced-a} has eigenvalues
$\pm\sqrt{\frac{2 \pi}{k}\mathcal{L}}$.  For the holonomy to be hyperbolic
with boost parameter $\lambda$, i.e.\ to
reproduce~\eqref{eq:hyperbolic-holonomy}, these eigenvalues must be
real and equal to $\pm\lambda/2$.  This requires
$\frac{2 \pi}{k}\mathcal{L}>0$ and
\beq\label{eq:L-lambda}
  \frac{2 \pi}{k}\mathcal{L} \;=\; \frac{\lambda^2}{4}
  \;=\; \frac{r_+^2}{4}\,,
\eeq
using $\lambda=r_+$ for the non-rotating case.  The same argument
applied to $\tilde a_\phi$ gives $\frac{2\pi}{k}\widetilde{\mathcal{L}}=r_+^2/4$.
Therefore
\beq\label{eq:aphi-BTZ}
  a_\phi \;=\; -\frac{r_+^2}{4}\;E \;+\; F\,,
  \qquad
  \tilde a_\phi \;=\; -\frac{r_+^2}{4}\;F \;+\; E\,.
\eeq

\medskip\noindent\textbf{Chirality fixes the temporal
  connection.}\quad
The Brown--Henneaux conditions in this chiral DS form set the
opposite lightcone components to zero, $a_-=0$ and
$\tilde a_+=0$.  Flatness then gives
$\partial_-a_+=0$ and $\partial_+\tilde a_-=0$, so that
$a=a_+(w^+)\,dw^+$ and $\tilde a=\tilde a_-(w^-)\,dw^-$, with
$w^\pm=\phi\pm t$.  Consequently,
$a_t=a_\phi$ and $\tilde a_t=-\tilde a_\phi$: the temporal
connection is not an independent parameter. The complete saddle is
therefore
\beq\label{eq:A-saddle}
  A \;=\; H\,d\rho \;+\;
  \Bigl(-\frac{r_+^2}{4}\;e^{-\rho}\,E
  \;+\; e^{\rho}\,F\Bigr)(d\phi + dt)\,,
\eeq
\beq\label{eq:Atilde-saddle}
  \tilde{A} \;=\; -H\,d\rho \;+\;
  \Bigl(-\frac{r_+^2}{4}\;e^{-\rho}\,F
  \;+\; e^{\rho}\,E\Bigr)(d\phi - dt)\,,
\eeq
with $A_\tau=-iA_\phi$ and $\tilde A_\tau=+i\tilde A_\phi$ in
Euclidean signature.  This is the complete saddle-point connection for
the microstate $|P\ket$, with no adjustable parameters: the holonomy
fixes $r_+$, the boundary conditions fix the solution, and
chirality fixes the temporal components.

\subsection{The Metric and the Euclidean
  Horizon}\label{sec:metric-horizon}

We now translate the saddle
connections~\eqref{eq:A-saddle}--\eqref{eq:Atilde-saddle} into the
metric language using the Chern-Simons / gravity
dictionary~\eqref{eq:CS-connections}:
\beq\label{eq:dreibein-from-CS}
  e^a\,J_a \;=\; \frac{\ell}{2}\bigl(A - \tilde{A}\bigr)\,,
  \qquad
  \omega^a\,J_a \;=\; \frac{1}{2}\bigl(A + \tilde{A}\bigr)\,,
\eeq
with $\ell=1$ and $g_{\mu\nu}=\eta_{ab}\,e^a_\mu e^b_\nu$.  The
dreibein, written as $e_\mu=e^a_\mu J_a$, is
\beq\label{eq:dreibein-explicit}
  e_\rho \;=\; H\,,
  \qquad
  e_\phi \;=\; \frac{f_+}{2}\,\bigl(F-E\bigr)\,,
  \qquad
  e_t \;=\; \frac{f_-}{2}\,\bigl(F+E\bigr)\,,
\eeq
where $f_\pm=e^{\rho}\pm\frac{r_+^2}{4}e^{-\rho}$.  This gives
$g_{\rho\rho}=1$, $g_{\phi\phi}=f_+^2$ and $g_{tt}=-f_-^2$, with no
cross terms.  The radial gauge~\eqref{eq:FG-gauge} is preserved by
constant shifts of $\rho$, which rescale the $E,F$ components of
$a_i$ and act as a constant Weyl rescaling of the boundary metric,
leaving both the holonomy and the Drinfel'd-Sokolov conditions
untouched.  The shift $\rho\to\rho+\log(r_+/2)$ places the centre of
the disc, where the Wilson line sits, at $\rho=0$, and turns $f_+$ and
$f_-$ into $r_+\cosh\rho$ and $r_+\sinh\rho$.  In terms of the standard
BTZ radial coordinate\footnote{The shifted coordinate $\rho$ covers
  $\rho\in[0,\infty)$, corresponding to $r\in[r_+,\infty)$.  In
  particular $\rho=0$ corresponds to $r=r_+$. The center of the
  spatial disc in the Chern-Simons formulation is the horizon in the
  metric language. In fact, this statement is true also for the rotating BTZ black hole. This is one reason why the discussion in this section can be generalized essentially trivially to that case as well. Note also that the horizon size is finite (``size" is a metric notion!), even though the source Wilson line is sitting at the origin in these coordinates.}
\beq\label{eq:rho-to-r}
  r \;=\; r_+\,\cosh\rho\,,
\eeq
the saddle is the Lorentzian BTZ metric
\beq\label{eq:BTZ-rho}
  ds^2 \;=\; -r_+^2\,\sinh^2\!\rho\;\,dt^2
  \;+\; d\rho^2
  \;+\; r_+^2\,\cosh^2\!\rho\;\,d\phi^2\,,
\eeq
equivalently
$ds^2=-(r^2-r_+^2)\,dt^2+dr^2/(r^2-r_+^2)+r^2\,d\phi^2$, which
matches~\eqref{eq:BTZ-metric} upon Wick rotation $t\to-i\tau$.  For future use, we also write the Euclidean metric in the form
\beq\label{eq:Euclidean-BTZ-rho}
  ds^2_E \;=\; r_+^2\,\sinh^2\!\rho\;\,d\tau^2
  \;+\; d\rho^2
  \;+\; r_+^2\,\cosh^2\!\rho\;\,d\phi^2\,.
\eeq
That the saddle is locally the BTZ metric is, of course,
unsurprising: the connection is flat with hyperbolic holonomy of boost
parameter $\lambda=r_+$, and since three-dimensional gravity has no
local degrees of freedom, the local geometry is completely determined
by the holonomy.  What the local analysis does not determine, and what
the path integral~\eqref{eq:PI-state-super} does, is the global
structure.

It is worth fixing the language of singularities here, since the gauge
field and the metric do not use the word in the same way.  In
Chern-Simons variables the statement is unambiguous:
by~\eqref{eq:CS-eom-source} the field strength has a delta function on
the Wilson line and vanishes elsewhere.  The metric
translation depends on the global structure.  If the Euclidean time
direction is compact with period $\beta$, then $\rho=0$ is a
codimension-2 locus at which the $\tau$-circle closes off, and unless
$\beta$ is tuned to $\frac{2 \pi}{r_+}$, there is a conical defect or excess there, with a
delta function in the curvature supported on it.  This is not what is
usually called a curvature singularity in the metric language, a
phrase reserved for diverging curvature scalars: here the geometry is
locally AdS$_3$ and its curvature scalars are finite everywhere except at the origin.  If instead the
Euclidean time direction is non-compact, as it is for the state prepared
by~\eqref{eq:PI-state-super}, then $\rho=0$ cannot be added as a
regular point at all, and the invariant statement is simply that the geometry is geodesically incomplete.  Therefore the context will clarify the meaning of the phrase,``singularity in curvature". With this in mind, we now examine the geometry near $\rho=0$.

\subsubsection{The Near-Horizon Geometry}\label{sec:near-horizon}

Near $\rho = 0$, the Euclidean
metric~\eqref{eq:Euclidean-BTZ-rho} reduces to
\beq\label{eq:near-horizon}
  ds^2_E \;\approx\; r_+^2\,\rho^2\;d\tau^2
  \;+\; d\rho^2
  \;+\; r_+^2\;d\phi^2\,.
\eeq
Two features are immediately apparent:
\begin{itemize}
\item The $\phi$-circle retains a \emph{finite} circumference
  $2\pi r_+$ at $\rho = 0$.  This is the \emph{horizon perimeter}.
\item The $(\rho,\tau)$-part of the metric,
  $d\rho^2 + r_+^2\rho^2\,d\tau^2$, is a two-dimensional metric in
  ``polar-like'' coordinates, with $\rho$ playing the role of the
  radial direction and $r_+\tau$ playing the role of the angle.
\end{itemize}
The nature of the geometry at $\rho = 0$ depends on what
range $\tau$ takes.  This is where the \emph{microstate} and the black hole
(thermal state) part ways.

\subsubsection{Thermal State vs.\ Microstate}\label{sec:thermal-vs-micro}

\medskip\noindent\textbf{The thermal (smooth) saddle.}\quad
If $\tau$ is periodically identified,
$\tau \sim \tau + \beta$, then the quantity $r_+\tau$ ranges over an
interval of length $r_+\beta$.  The $(\rho,\tau)$-geometry is
smooth at $\rho = 0$ if and only if $r_+\beta = 2\pi$, i.e.\
$\beta = 2\pi/r_+$.  This is the standard Gibbons--Hawking argument:
smoothness of the Euclidean cigar geometry fixes the
temperature.  The resulting geometry is the smooth Euclidean
BTZ black hole, topologically a solid torus, with the thermal
$\tau$-circle contractible at the tip $\rho = 0$.

\medskip\noindent\textbf{The microstate saddle.}\quad
For the individual microstate $|P\ket$, the situation is
qualitatively different.  The state is prepared by the path integral
on the half-cylinder $D^2 \times (-\infty,\,0]$--the
Euclidean time $\tau$ ranges over $(-\infty,\,0]$ and is non-compact.  There is no periodic identification of $\tau$,
and therefore the Gibbons--Hawking smoothness argument does not
apply.

The $(\rho,\tau)$-geometry near $\rho = 0$ is ``polar coordinates''
with an angular direction that ranges over $(-\infty,0]$ rather than
$[0,2\pi)$.  This is not a smooth point, not a cone, but a
degenerate surface.  In the Chern-Simons language, the origin of this
degeneracy is clear: there is a Wilson line source at $\rho = 0$.
The field strength has a delta-function singularity
(essentially eq.~\eqref{eq:CS-eom-source}, but with a hyperbolic holonomy source), 
localised at the centre of the disc, which is the horizon $r = r_+$
in the metric variables~\eqref{eq:rho-to-r}.  In the Chern-Simons
formulation this source is simply part of the classical equations of
motion--the connection is smooth everywhere away from it.  But in
the metric language, the source shows up as a singularity at the
horizon.

To summarise: in the microstate saddle, the Euclidean geometry
covers the region $r \geq r_+$ (equivalently $\rho \geq 0$), with the
$\phi$-circle retaining a finite horizon perimeter $2\pi r_+$, but
with no smooth contraction of the $\tau$-direction.  The ``horizon''
is the location of the Wilson line source--a singularity in the
curvature, not a smooth interior point.

This is exactly the Euclidean geometry described in
Section~4.2 of~\cite{CKPSP}: the centre of the disc
carries a Wilson line in the Chern-Simons language, which translates
to a finite ``horizon area'' (here, perimeter) in the metric language,
with no smoothness condition because the Euclidean time direction need
not be periodic.\footnote{In the language
  of~\cite{CKPSP}, the Wilson line plays the role of the
  inner boundary, and the presence of a source there is the gauge-theoretic avatar of the absence of
  a smooth horizon for an individual microstate.}

\subsection{Lorentzian Signature: Singular
  Horizon}\label{sec:Lorentzian-horizon}

We now Wick-rotate back to Lorentzian signature and examine the
horizon of the individual microstate.

The Lorentzian metric~\eqref{eq:BTZ-rho} describes the
\emph{exterior} of the BTZ black hole, valid for $\rho > 0$ (i.e.\
$r > r_+$).  Near the horizon $\rho \to 0^+$, the metric
reduces to
\beq\label{eq:Rindler}
  ds^2 \;\approx\; -r_+^2\,\rho^2\;dt^2
  \;+\; d\rho^2
  \;+\; r_+^2\;d\phi^2\,.
\eeq
The $(\rho,t)$-part is simply Rindler space -- the
near-horizon approximation familiar from any non-extremal
black hole.  Whether the Rindler horizon at $\rho = 0$ is a smooth,
extendable surface or a genuine singularity depends on the
\emph{quantum state}.

\medskip\noindent\textbf{The modular thermal state: smooth Kruskal
  extension.}\quad
A {\em thermal state} by itself need not have a smooth horizon: the KMS
condition can be imposed for any thermal ensemble.  What is special
about the black-hole state is the density of microstates.  As we elaborate
in Section~\ref{sec:thermal}, when the thermal sum is weighted by the
density dictated by modular invariance, its dominant saddle satisfies
$\beta r_+=2\pi$.  The Euclidean thermal cycle is then smoothly
contractible, and after analytic continuation the geometry extends
through $\rho=0$ to the full two-sided BTZ spacetime with a regular
Kruskal horizon.  Thus smoothness is a property of the modular thermal
ensemble, rather than a consequence of KMS periodicity alone.

\medskip\noindent\textbf{The individual microstate: singular
  horizon.}\quad
The microstate $|P\ket$ is a {\em pure state}, prepared by the
Euclidean path integral on the non-compact half-cylinder.  Its
Euclidean preparation involves a Wilson line source at $\rho = 0$--a
delta-function in the gauge field curvature--and no periodic
identification of Euclidean time.  As a consequence:
\begin{enumerate}
\item The state does not satisfy KMS periodicity, nor does it have a modular density -- it is a single eigenstate. 
\item The Wilson line source at $\rho = 0$ is a
  genuine singularity that sources the CS field strength, not a coordinate
  artifact.  Concretely, the CS field strength 
  has a delta-function (source) at $\rho = 0$.  This is a statement about the
  spatial slice--the punctured disc at any fixed time--and is
  therefore the same in both Euclidean and Lorentzian
  signatures.\footnote{The equation of
    motion~\eqref{eq:CS-eom-source} is a constraint on the spatial
    slice at any fixed time $t$ (or $\tau$).  Since the spatial disc
    is the same regardless of whether the time direction is Lorentzian
    or Euclidean, the delta-function source at $\rho = 0$ carries over
    unchanged.}
\item The Lorentzian geometry of the microstate
  terminates at $\rho = 0$ (i.e.\ $r = r_+$) with a
  finite-perimeter.  The horizon is not a smooth null surface through which the geometry can be extended\footnote{There is no contradiction with the fact that the exterior metric
\eqref{eq:BTZ-rho}, considered by itself, admits the usual smooth BTZ
Kruskal completion. That is a geometry designed to capture thermality. In the microstate saddle, $\rho=0$ is the
Wilson-line puncture, with a delta-function source for the Chern--Simons
field strength and non-trivial holonomy around the linking $\phi$-cycle. There is no natural notion of thermality, but there is a natural notion of a source sitting at $r=r_+$. 
The usual two-sided BTZ completion instead removes this source and changes
the global completion of the spatial slice by extending through the
would-be horizon to a second exterior.  The same hyperbolic $\phi$-holonomy
is then supported by the non-contractible cycle of the source-free BTZ
geometry.  So we will view the existence of a smooth source-free completion as the answer to a different problem.}.
\end{enumerate}

\subsubsection{The Nature of the Singular Horizon}\label{sec:singularity-nature}

It is important to be precise about what kind of singularity the
microstate horizon is.  The Riemann tensor of the BTZ metric is that
of locally AdS$_3$ everywhere--there is no curvature singularity in
the usual Riemannian sense away from $\rho = 0$.  The singularity is
of a different character:
\begin{itemize}
\item In the \textbf{Chern-Simons formulation}, the connection is
  smooth and flat on the punctured disc $\rho > 0$, with a
  delta-function curvature source at $\rho = 0$.  The gauge field
  description is completely regular away from the source.

\item In the \textbf{metric formulation}, the delta-function in
  $F_{ij}$ translates to a delta-function contribution to the
  curvature at $\rho = 0$\footnote{This translation assumes a compact thermal cycle. More generally, it can be viewed as being geodesically incomplete.}.  The resulting singularity is analogous to
  the conical singularity at the tip of a conical defect (which
  also has distributional curvature supported at a point), but
  occurring now at a surface of finite perimeter.
\end{itemize}

An instructive comparison is with the sub-threshold (conical defect)
case.  For the conical defect microstate constructed in
Section~\ref{sec:microstates}, the saddle metric
is~\eqref{eq:conical-metric}, with a conical singularity at $r = 0$
where the $\phi$-circle collapses.  The Wilson line source sits at
the tip $r = 0$, producing a delta-function curvature that
encodes the deficit angle.  For the super-threshold (BTZ) microstate,
the situation is qualitatively the same--a Wilson line source at
the origin of the disc, producing a delta-function in the
curvature--except that the $\phi$-circle no longer collapses at the
source\footnote{The word curvature can be read to mean either the CS field strength or the metric curvature in this sentence, but the collapse of the $\phi$-circle is a metric reading.}.  Instead, it retains a finite circumference $2\pi r_+$, which
is the horizon perimeter.

It is crucial to not forget that the singularity of the microstate is an artifact of the semi-classical (large-$k$) limit that we have taken on the quantum state. At any finite value of $k$, the state constructed by our Wilson line path integral is a perfectly well-defined Hilbert space state. But in the semi-classical limit, we find gauge fields with singular sources or metrics with singular horizons.

\subsection{Smooth Horizon as an   Ensemble of States}\label{sec:smooth-vs-singular}

In both cases (black holes as well as microstates), the Chern-Simons holonomy around the spatial circle
can be taken to be in the same hyperbolic conjugacy class, with the same boost
parameter $\lambda = r_+$.  The local geometry is the same--both are
locally the BTZ metric. 

In the BTZ black hole, Euclidean geometry is a solid torus $D^2 \times S^1_{2 \pi}$
  with periodic $\tau$, with period $\beta$. (Note that the disc here is not the spatial disc.)
  There is no Wilson line: the disc is unpunctured.
The saddle is the smooth Euclidean BTZ with the thermal
  $\tau$-circle contractible at $\rho = 0$.  Smoothness condition
  fixes $\beta = 2\pi/r_+$.
In the Lorentzian geometry, this continues to the Hartle--Hawking state on the full
  two-sided BTZ spacetime. This is the Kruskal horizon. The holonomy is hyperbolic, boost parameter $\lambda = r_+$,
fixed by the temperature (i.e.\ by the inverse temperature
  $\beta$).

We emphasise that the holonomy is determined by different physical
inputs in the two cases.  For the microstate, it is the state label $s$ (equivalently, the conformal weight
$h_P$) that fixes $\lambda = 2s/k$.  For the thermal state, it is the
temperature $T = 1/\beta$ that fixes $\lambda$ via the
smoothness condition $\beta = 2\pi/\lambda$.  In the microstate, we
specify the state, in the thermal case, we specify the
temperature.  The smooth horizon emerges only in the latter. 

The situation when there is rotation, is similar. To specify the microstate, we now have to specify two holonomies, while to specify the ensemble we have to specify two potentials (the temperature and the angular velocity, or equivalently, left and right temperatures). The black hole can be understood
as a grand-canonical partition function worth of such microstates. We will discuss this more in the next section. This picture provides a concrete, controlled realization of the idea that smooth horizons are emergent, coarse-grained features of thermal ensembles. Usually, we only have access to Gibbons-Hawking-like calculations which do not tell us what the individual microstates look like. Our point here is that one can construct these quantum microstates: considered individually, they have singular horizons in the semi-classical limit.

But we can do more: the thermal partition function is in fact a weighted sum over the microstates, with the (approximate) density of states dictated by modular invariance. Studying the thermal saddles and fluctuations of this partition function is instructive, and that is what we turn to, next.

\section{Smoothness and Fluctuations}\label{sec:thermal}

A thermal ensemble of microstates is useful as a statistical system if we have some understanding about {\em which} of these microstates are present in a given theory. In most situations that are of interest in statistical mechanics, particularly when strong chaos and thermalization are present, we do not know an exact answer to this question. But approximate knowledge can go a long way\footnote{Note for example, that we have a decent understanding of metals, despite the fact that they are almost impossibly complicated systems.}. In the case of holographic CFTs, in the high temperature limit, such an approximation becomes available to us thanks to modular invariance. In this regime, the vacuum in the dual channel is the dominant contribution to the partition function. Remarkably, the dual channel vacuum has a re-writing as a continuum sum over super-threshold (i.e., heavy) primaries in terms of a function called the modular $S$-kernel. This means that in the high temperature limit, the partition function is approximately a continuum sum over explicitly known microstates. This allows us to do and interpret some simple calculations and connect them to known physics. This is the goal of this section. The leading saddle will reproduce well-known results associated to the BTZ black hole, but we will also compute the fluctuations. This is a trivial computation for us, but its form will match the (somewhat complicated) ``fluctuations of the area" type results in general relativity \cite{Parikh:2024area, Ciambelli:2025area}.

A key fact about the above set up, is that it does {\em not} rely on the large-$c$ limit. It is instead reliant on the Cardy regime for its validity, and therefore in principle, allows us to go beyond the leading large-$c$ limit. In this paper our goals will be more modest: we want to connect with (semi-)classical gravitational expectations and therefore we will be working in the large-$c$ (large-$k$) limit. 

The perspective of this section is that 
we fix the temperature (and angular chemical potential), and the
holonomy is determined dynamically as a saddle point of the
ensemble integral.  As we will show, the saddle automatically
produces the smooth Euclidean BTZ--the smoothness of the thermal
cycle is not imposed by hand but emerges as a saddle-point
condition.  Moreover, the Gaussian fluctuation around the saddle at
finite $k$ yields a variance $\text{Var}(S) = S$ for the
Bekenstein--Hawking entropy, reproducing the BTZ version of
``quantum fluctuations in the area'' expected from perturbative
general relativity.

\subsection{From the Microstate to the Ensemble}\label{sec:micro-to-ensemble}

Our discussion in the previous section was framed in the non-rotating BTZ setting, for simplicity of exposition. But the discussion is most naturally set up for a general
rotating black hole, and since carrying the rotation along costs
nothing, we do so throughout this section.  The non-rotating statements
are recovered by setting $\tau_1=0$ at the end.

The dictionary between the modular parameter $\tau=\tau_1+i\tau_2$ of
the boundary torus and the thermodynamic variables is,
\bea\label{eq:tau-dictionary}
  q \;=\; e^{2\pi i\tau}\,,
  &\qquad&
  \bar q \;=\; e^{-2\pi i\bar\tau}\,,
  \non\\[4pt]
  \beta \;=\; 2\pi\tau_2\,,
  &\qquad&
  \theta \;=\; 2\pi\tau_1\,,
\eea
with $\beta$ the inverse temperature and $\theta$ the potential
conjugate to the angular momentum $J=L_0-\bar L_0$.  Since the two
chiral sectors will decouple, it is convenient to trade $\tau$ and
$\bar\tau$ for the left- and right-moving inverse temperatures
\beq\label{eq:betaLR}
  \beta_L \;=\; -2\pi i\,\tau \;=\; \beta-i\theta\,,
  \qquad
  \beta_R \;=\; 2\pi i\,\bar\tau \;=\; \beta+i\theta\,,
\eeq
so that $q=e^{-\beta_L}$, $\bar q=e^{-\beta_R}$ and
$\beta=\frac12(\beta_L+\beta_R)$.  For a Lorentzian rotating black hole
$\beta_L$ and $\beta_R$ are separately real and positive, which
corresponds to imaginary $\tau_1$.  We will not impose this, and treat
$\tau$ and $\bar\tau$ as independent complex parameters throughout.

The grand-canonical partition function on the boundary torus is
\beq\label{eq:Z-torus}
  Z(\tau,\bar\tau)
  \;=\;
  \Tr\; q^{L_0-c/24}\;\bar q^{\bar L_0-c/24}\, ,
\eeq
and the object of interest for us is ``the BTZ partition function". This is the $S$-modular image of the vacuum character,
\beq\label{eq:Z-BTZ}
  Z_{\rm BTZ}(\tau,\bar\tau)
  \;=\;
  \chi_{\rm vac}\!\left(-\frac{1}{\tau}\right)
  \bar\chi_{\rm vac}\!\left(-\frac{1}{\bar\tau}\right),
\eeq
which dominates $Z$ at high temperature.  This factorises into a
holomorphic and an antiholomorphic piece, so we present every
computation in the holomorphic sector and obtain its counterpart by
$\tau\to\bar\tau$, $P\to\bar P$, $\beta_L\to\beta_R$.  When the black
hole rotates the two sectors sit at different saddle points, and this
is how the two horizon radii $r_\pm$ enter.

\subsection{The $S$-Kernel as a Density of Primaries}
\label{sec:S-kernel-density}

The modular transform of the vacuum is an integral over
non-degenerate Virasoro characters,
\beq\label{eq:S-decomposition}
  \chi_{\rm vac}\!\left(-\frac{1}{\tau}\right)
  \;=\;
  \int_0^\infty dP\;\; S_{0P}\;\;\chi_P(\tau)\,,
  \qquad
  \chi_P(\tau) \;=\; \frac{q^{P^2}}{\eta(\tau)}\, ,
\eeq
where $\chi_P$ is the character of the Verma module built on the
primary of weight $h_P=\frac{c-1}{24}+P^2$ of eq.~\eqref{eq:h-P}, the
factor $1/\eta(\tau)$ accounting for its Virasoro descendants.  The
kernel is \cite{CKPSP,CEZ1,CEZ2}
\beq\label{eq:S-kernel-explicit}
  S_{0P} \;=\; 4\sqrt{2}\;\sinh\bigl(2\pi bP\bigr)\;
  \sinh\!\left(\frac{2\pi P}{b}\right),
\eeq
with $b$ the Liouville parameter of eq.~\eqref{eq:h-P}, $b^{-2}=k-2$.
In the gravity regime $k\gg1$ we have $b\ll1$, $Q\approx1/b$ and
$c\approx6k$.

$S_{0P}$ is the density of Virasoro primaries at momentum $P$ dictated
by modular invariance.  Extracting the leading exponential from each
$\sinh$ gives the exact rewriting
\beq\label{eq:S0P-asymp}
  \log S_{0P}
  \;=\;
  2\pi\,Q\,P \;+\; \log\sqrt{2}
  \;+\;
  \log\Bigl[\bigl(1-e^{-4\pi bP}\bigr)
  \bigl(1-e^{-4\pi P/b}\bigr)\Bigr],
\eeq
which reduces to $\log S_{0P}\approx2\pi QP$ once $P$ is large compared
to both $b$ and $1/b$.  Since $h_P\approx P^2$ for such heavy states
and $Q^2=\frac{c-1}{6}$, this is
\beq\label{eq:Cardy-check}
  \log S_{0P} \;\approx\; 2\pi\sqrt{\frac{c-1}{6}\,h_P}\, ,
\eeq
the Cardy growth of the primary degeneracy, with the shift $c\to c-1$
that separates primaries from descendants already built in.  This is
the sense in which the kernel supplies the microstates: it counts the
heavy primaries of Section~\ref{sec:super-threshold}, each of which we
have constructed as a Wilson-line state.

Assembling both chiralities,
\beq\label{eq:Z-BTZ-full}
  Z_{\rm BTZ}(\tau,\bar\tau)
  \;=\;
  \frac{1}{\eta(\tau)\,\bar\eta(\bar\tau)}
  \int_0^\infty\!\!dP \int_0^\infty\!\!d\bar P\;\;
  S_{0P}\;S_{0\bar P}\;\;
  q^{P^2}\;\bar q^{\bar P^2}\, .
\eeq

\subsection{Saddle-Point Evaluation at Large $k$}\label{sec:saddle}

All the $P$-dependence of \eqref{eq:Z-BTZ-full} sits in the holomorphic
spectral integral
\beq\label{eq:I-def}
  I(\tau) \;\equiv\; \int_0^\infty dP\;\;S_{0P}\;\;
  e^{2\pi i\tau\,P^2}
  \;=\;
  \int_0^\infty dP\;\;S_{0P}\;\;e^{-\beta_L P^2}\, ,
\eeq
which we evaluate by steepest descent at large $k$.  The exponent
\beq\label{eq:f-def}
  f(P) \;=\; \log S_{0P} \;+\; 2\pi i\tau\,P^2
\eeq
is stationary at $f'(P_*)=0$.  Using
\beq\label{eq:dlog-S}
  \frac{\de}{\de P}\,\log S_{0P}
  \;=\;
  2\pi b\,\coth\bigl(2\pi bP\bigr)
  \;+\;
  \frac{2\pi}{b}\,\coth\!\left(\frac{2\pi P}{b}\right),
\eeq
the saddle-point condition reads
\beq\label{eq:saddle-full}
  2\pi b\,\coth\bigl(2\pi bP_*\bigr)
  \;+\;
  \frac{2\pi}{b}\,\coth\!\left(\frac{2\pi P_*}{b}\right)
  \;=\;
  -4\pi i\tau\,P_* \;=\; 2\,\beta_L\,P_*\, ,
\eeq
which is exact at any $k$.

The two arguments appearing here have a transparent meaning.
With the exact relation $P=sb$ from
eq.~\eqref{eq:Liouville-P} and the leading semiclassical relation
$\lambda=2s/k+O(1/k)$ from eq.~\eqref{eq:s-lambda},
\beq\label{eq:saddle-arguments}
  \frac{P_*}{b} \;=\; s_*\, ,
  \qquad
  2b\,P_* \;=\; \frac{2s_*}{k-2}
  \;=\; \lambda_L\bigl(1+O(1/k)\bigr)\, ,
\eeq
so the second $\coth$ in \eqref{eq:saddle-full} has argument $2\pi s_*$
and the first has argument
$\pi\lambda_L\bigl(1+O(1/k)\bigr)$.  For a macroscopic black
hole $s_*\gg1$, so the second $\coth$ is unity up to
$O(e^{-4\pi s_*})$, while $\lambda_L$ is of order one, so the first
term is $O(b)$ against the $O(1/b)$ of the second and is a relative
correction of order $1/k$.  This leads to,
\beq\label{eq:saddle-leading}
  \frac{2\pi}{b} \;=\; 2\,\beta_L\,P_*
  \qquad\Longrightarrow\qquad
  P_* \;=\; \frac{\pi}{b\,\beta_L} \;=\; \frac{i}{2b\tau}\, ,
\eeq
and in the antiholomorphic sector
$\bar P_*=\pi/(b\beta_R)=-i/(2b\bar\tau)$.\footnote{For general complex
$\tau$ the saddle lies off the real axis, and the contour is deformed
into the complex $P$-plane to pass through it.  This is standard
steepest descent, and the Gaussian analysis of
Section~\ref{sec:fluctuations} proceeds on the deformed contour exactly
as it would on the real axis.}

By \eqref{eq:saddle-arguments}, the boost parameters of the two spatial
holonomies at the saddle are therefore, at leading order in $1/k$,
\beq\label{eq:lambda-from-saddle}
  \lambda_L \;=\; 2b\,P_* \;=\; \frac{2\pi}{\beta_L}
  \;=\; \frac{i}{\tau}\, ,
  \qquad
  \lambda_R \;=\; 2b\,\bar P_* \;=\; \frac{2\pi}{\beta_R}
  \;=\; -\frac{i}{\bar\tau}\, .
\eeq
The horizon radii $r_\pm=\frac12(\lambda_L\pm\lambda_R)$ of
eq.~\eqref{eq:rotating-match} (we set AdS length $\ell=1$ in all of this), are
thus fixed by the temperature and the angular potential.  In the non-rotating limit
$\theta=0$ we have $\beta_L=\beta_R=\beta$, $r_-=0$ and
$\lambda=r_+=2\pi/\beta$, while $s_*=P_*/b=\pi(k-2)/\beta$, which at
leading order in large $k$ is the identification $s=kr_+/2$ of
eq.~\eqref{eq:s-lambda}.

\subsection{Smooth Horizon from the Saddle}
\label{sec:smooth-from-saddle}

The Euclidean geometry associated with the ensemble is a solid torus:
the boundary torus of modular parameter $\tau$, filled in so that the
thermal cycle is contractible.  The geometry is smooth where that cycle
closes off precisely when the Chern-Simons holonomy around it is
trivial (i.e., central in each of the two gauge groups).  For the
rotating BTZ solution this is the pair of conditions
\beq\label{eq:smoothness-condition}
  \lambda_L\,\beta_L \;=\; 2\pi\, ,
  \qquad
  \lambda_R\,\beta_R \;=\; 2\pi\, ,
\eeq
equivalently $\beta_{L,R}=2\pi/(r_+\pm r_-)$.  This is exactly what the
saddle delivers in eq.~\eqref{eq:lambda-from-saddle}, in both chiral
sectors and for every $\tau$.

In the metric language, the pair \eqref{eq:smoothness-condition} is the
pair of conditions that one usually imposes separately.  Written in
terms of $\beta$ and $\theta$, they are
\beq\label{eq:smoothness-metric}
  \beta \;=\; \frac{2\pi r_+}{r_+^2-r_-^2}\, ,
  \qquad
  i\,\theta \;=\; \beta\,\Omega_H\, ,
  \qquad
  \Omega_H \;=\; \frac{r_-}{r_+}\, ,
\eeq
or equivalently $\beta_{L,R}=\beta\,(1\mp\Omega_H)$.  The first is the
absence of a conical defect at the outer horizon, which is the
statement that $1/\beta$ is the Hawking temperature.  The second fixes
the twist of the thermal cycle: the cycle that closes off is the one
generated by $\de_t+\Omega_H\,\de_\phi$, the Killing vector that becomes
null on the horizon, and not by $\de_t$.  The factor of $i$ is the
remark made below eq.~\eqref{eq:betaLR}, that $\tau_1$ is imaginary for
a Lorentzian rotating black hole, so both sides of the second relation
are real.  Only the first condition survives at $r_-=0$, where
$\theta=0$ and $\beta=2\pi/r_+$.  In the Chern-Simons language, the two conditions turn into parallel statements about the triviality of the holonomy around the contractible cycle, imposed in each of the two gauge groups.

It is worth being explicit about what is and is not being assumed here.
The microstate of Section~\ref{sec:super-threshold} is specified through
the pair of representation labels $s_L$ and $s_R$, and its two spatial
holonomies are $\lambda_{L,R}=2s_{L,R}/k$ by
eq.~\eqref{eq:rotating-match}.  In the semi-classical analysis of
Section~\ref{sec:semiclassical} the Euclidean time direction was
non-compact, so that no smoothness condition was in effect: the geometry
had a singular horizon.   Here the temperature
and the angular potential are specified instead, and the holonomies are
obtained as saddles.  In each chiral sector separately, the holonomy comes from the
competition in \eqref{eq:I-def} between the growth of the density of
primaries, $S_{0P}\sim e^{2\pi QP}$, and the Boltzmann suppression
$e^{-\beta_L P^2}$.  The data is a pair of numbers in both cases, but the same two hyperbolic conjugacy
classes are fixed by different physics:
\bea\label{eq:holonomy-origins}
  \hbox{Microstate:} &\qquad&
  \lambda_{L,R} \;=\; \frac{2s_{L,R}}{k}\, ,
  \qquad \hbox{fixed by the representation labels,}
  \non\\[4pt]
  \hbox{Ensemble:} &\qquad&
  \lambda_{L,R} \;=\; \frac{2\pi}{\beta_{L,R}}\, ,
  \qquad \hbox{fixed by the saddle point.}
\eea
The smoothness of the thermal cycle is a thermodynamic/statistical statement
about the second, not a geometric input.  This is the sense in which
the smooth horizon is a property of the ensemble and not of the
individual states that comprise it.

\subsection{Lorentzian Smoothness, TFD and the Modular Ensemble}
\label{sec:modular-ensemble}
 
Because the answer is still the smooth horizon, it is easy to mistake that the saddle calculation is a trivial re-writing of the familiar Gibbons-Hawking calculation. So let us emphasize that it is quite different: it provides a statistical mechanics answer to black hole thermodynamics, while Gibbons-Hawking provided a geometric answer. The saddle balances the growth of the density of primaries against the
Boltzmann factor, and the answer $\lambda_{L,R}=2\pi/\beta_{L,R}$ came
out of the specific exponential growth $S_{0P}\sim e^{2\pi QP}$ of the
modular kernel.  To clarify this, it is worth asking what would happen if that growth
were different.  Modular invariance fixes the kernel uniquely of course, but tuning it ``by hand" is useful as a diagnostic: a way of identifying which feature of the spectrum the
smoothness of the horizon is sensitive to.
 
Replace the kernel in \eqref{eq:I-def} by a density of heavy primaries
with a rescaled Cardy growth,
\beq\label{eq:deformed-density}
  I_\gamma(\tau) \;=\; \int_0^\infty dP\;\; \rho_\gamma(P)\;\;
  e^{-\beta_L P^2}\, ,
  \qquad
  \rho_\gamma(P) \;=\; e^{2\pi\gamma QP}\, ,
  \qquad
  \gamma>0\, ,
\eeq
so that $\gamma=1$ is the modular value.  The saddle-point condition is
now $2\pi\gamma Q=2\beta_L P_*$, so that $P_*=\pi\gamma Q/\beta_L$, and
the boost parameter of the spatial holonomy at the saddle is
$\lambda_L=2bP_*=2\pi\gamma/\beta_L$ at leading order in $1/k$, with the
same statement in the antiholomorphic sector.  In place of
\eqref{eq:smoothness-condition} one therefore finds
\beq\label{eq:deformed-smoothness}
  \lambda_L\,\beta_L \;=\; 2\pi\gamma\, ,
  \qquad
  \lambda_R\,\beta_R \;=\; 2\pi\gamma\, .
\eeq
The holonomy around the contractible cycle is non-trivial unless
$\gamma=1$.  In the metric language, by the near-horizon form
\eqref{eq:near-horizon}, the $(\rho,\tau)$ plane now closes off with
total angle $2\pi\gamma$ instead of $2\pi$: the filled solid torus
carries a conical defect at its core for $\gamma<1$ and a conical excess
for $\gamma>1$, located at the horizon.
 
It is useful to say what $\gamma$ measures.  Since $h_P\approx P^2$ for
heavy states and $Q^2=\frac{c-1}{6}$, the growth in
\eqref{eq:deformed-density} can be written as in \eqref{eq:Cardy-check}
with $c$ replaced by an effective value,
\beq\label{eq:c-effective}
  \log\rho_\gamma \;\approx\; 2\pi\sqrt{\frac{c_{\rm eff}-1}{6}\,h_P}\, ,
  \qquad
  c_{\rm eff}-1 \;=\; \gamma^2\,(c-1)\, .
\eeq
A deformed density is thus one whose Cardy growth is governed by a
central charge different from the $c$ that fixes the Newton constant
through $c \approx 6k$.  This mismatch shows up in a second place as well.
Evaluating \eqref{eq:deformed-density} at its saddle by the steps of
Section~\ref{sec:on-shell-action} gives
$\log I_\gamma=\pi^2\gamma^2Q^2/\beta_L$, so the entropy is $\gamma^2$
times its value at $\gamma=1$, while the horizon radius
$r_+=\frac12(\lambda_L+\lambda_R)$ is by \eqref{eq:deformed-smoothness}
only $\gamma$ times its value at $\gamma=1$.  At leading order in $c$ the two do not
agree,
\beq\label{eq:S-gamma-area}
  S \;=\; \gamma\;\frac{A}{4G_N}\, ,
\eeq
which reduces to \eqref{eq:S-BH-total} at $\gamma=1$.  The
Bekenstein-Hawking relation and the smoothness of the thermal cycle fail
together, and both hold only at the modular value.
 
We can now state the Lorentzian version.  For any density of states one
may form the thermal density matrix at inverse temperature $\beta$ and
purify it in a doubled Hilbert space as the thermofield double,
\beq\label{eq:TFD}
  |{\rm TFD}\rangle \;\propto\; \sum_n e^{-\beta E_n/2}\,
  |n\rangle_L\otimes|n\rangle_R\, ,
\eeq
the sum running over the energy eigenstates of the boundary theory, so
that the density of states enters through the multiplicities.  This
construction is kinematical: it is available for any spectrum and any
temperature, and tracing out one factor returns the thermal state on the
other.  In the bulk, the Euclidean preparation of \eqref{eq:TFD} is the
path integral on half of the filled torus, cut along $\tau=0$ and
$\tau=\beta/2$, and the $t=0$ slice of the resulting Lorentzian geometry
is the Einstein-Rosen bridge joining the two boundaries \cite{Israel, Maldacena}.  The
bifurcation surface of that geometry is the core of the solid torus, the
same locus that carries the conical defect above.  It is a regular
bifurcate Killing horizon, through which the geometry continues to the
full Kruskal extension, exactly when the Euclidean tip is smooth\footnote{Another way to think about this is that Euclidean smoothness and Lorentzian smoothness are two faces of the same {\em analytic} condition here.}.  By
\eqref{eq:deformed-smoothness} this happens only at $\gamma=1$.
 
Two statements that are often used interchangeably therefore come apart.
The thermofield double is available for any spectrum.  But not all TFDs correspond to an eternal
Lorentzian black hole with a smooth horizon.  When one writes the
eternal black hole as a thermofield double and treats the horizon as a
regular surface through which the geometry extends, one has implicitly
assumed that the weights in \eqref{eq:TFD} are the modular ones.  We
refer to the ensemble with that density of states as the {\em modular ensemble}.  It is
worth putting this beside the result of
Section~\ref{sec:semiclassical}.  There, an individual microstate was
found to have a singular horizon, its holonomy being fixed by the
representation label rather than by a saddle point.  Here we see that
averaging is not by itself the remedy: a thermal average taken with a
non-modular density of primaries is singular at the horizon as well.
Smoothness requires a regime where the modular ensemble is a good approximation to the spectrum. For BTZ, this is the Cardy regime.

\subsection{On-Shell Action and the Bekenstein-Hawking Entropy}
\label{sec:on-shell-action}
 
Before turning to fluctuations, we present the value of the integral at
the saddle.  By \eqref{eq:saddle-arguments} the two hyperbolic
cotangents in \eqref{eq:saddle-full} have arguments
$\pi\lambda_L\bigl(1+O(1/k)\bigr)$ and
$2\pi s_*$ there, and the last term of \eqref{eq:S0P-asymp} involves
the same two combinations.  Both are large in the regime of interest,
so up to exponentially small corrections the cotangents equal unity
and that term is negligible.  The saddle-point condition then gives
$2\pi Q=2\beta_L P_*$, so that $P_*=\pi Q/\beta_L$, which agrees with
\eqref{eq:saddle-leading} up to a relative correction of order $1/k$,
and $\log S_{0P_*}=2\pi QP_*+O(1)$.  Using $2\pi i\tau=-\beta_L$,
\beq\label{eq:on-shell-eval}
  \log I \;\approx\; f(P_*)
  \;=\; 2\pi\,Q\,P_* \;-\; \beta_L\,P_*^2
  \;=\; \frac{2\pi^2Q^2}{\beta_L} \;-\; \frac{\pi^2Q^2}{\beta_L}
  \;=\; \frac{\pi^2Q^2}{\beta_L}
  \;=\; \frac{\pi^2(c-1)}{6\,\beta_L}\, ,
\eeq
where the last step uses $c=1+6Q^2$ from \eqref{eq:c-Liouville}.  This
is the contribution of the primaries alone, and it is the Cardy free
energy of a theory of central charge $c-1$, the integrated form of the
primary counting \eqref{eq:Cardy-check}.
 
The $\eta$-function prefactors in \eqref{eq:Z-BTZ-full} carry the
Virasoro descendants.  With $\tau=i\beta_L/2\pi$, the modular
transformation $\eta(-1/\tau)=\sqrt{-i\tau}\;\eta(\tau)$ gives
\beq\label{eq:eta-asymptotics}
  \eta(\tau) \;=\; \sqrt{\frac{2\pi}{\beta_L}}\;\,
  e^{-\pi^2/(6\beta_L)}\;
  \Bigl(1+O\bigl(e^{-4\pi^2/\beta_L}\bigr)\Bigr)\, ,
\eeq
so that $\log\bigl(1/\eta(\tau)\bigr)=\pi^2/(6\beta_L)$ up to a term of
order $\log\beta_L$.  This is the Cardy free energy at $c=1$, and
adding the two pieces \cite{CKPSP},
\beq\label{eq:free-energy-chiral}
  \log I \;+\; \log\frac{1}{\eta(\tau)}
  \;=\; \frac{\pi^2(c-1)}{6\,\beta_L}
  \;+\; \frac{\pi^2}{6\,\beta_L}
  \;=\; \frac{\pi^2 c}{6\,\beta_L}\, ,
\eeq
which realizes the split $(c-1)+1=c$ of
Section~\ref{sec:super-threshold} quantitatively.  The antiholomorphic
sector gives the same expression with $\beta_L\to\beta_R$, so
\beq\label{eq:free-energy}
  \log Z_{\rm BTZ}
  \;\approx\;
  \frac{\pi^2 c}{6}\left(\frac{1}{\beta_L}+\frac{1}{\beta_R}\right),
\eeq
the high-temperature free energy of a CFT of central charge $c$.  The
entropy of each chiral sector follows from
\eqref{eq:free-energy-chiral} by the usual thermodynamic relation,
\beq\label{eq:S-BH-chiral}
  S_L \;=\;
  \left(1-\beta_L\,\frac{\de}{\de\beta_L}\right)
  \left(\log I+\log\frac{1}{\eta}\right)
  \;=\; \frac{\pi^2 c}{3\,\beta_L}
  \;\approx\; 2\pi\sqrt{\frac{c\,h_*}{6}}\, ,
  \qquad
  h_* \;=\; P_*^2\, ,
\eeq
which is the Cardy formula \cite{Carlip}, and similarly for $S_R$.
Using $\lambda_{L,R}=2\pi/\beta_{L,R}$ and
$r_+=\frac12(\lambda_L+\lambda_R)$, the total entropy is
\beq\label{eq:S-BH-total}
  S \;=\; S_L+S_R
  \;=\; \frac{\pi c}{6}\bigl(\lambda_L+\lambda_R\bigr)
  \;=\; \frac{\pi\,c\,r_+}{3}
  \;=\; \frac{2\pi r_+}{4G_N}\, ,
\eeq
the Bekenstein-Hawking entropy of the rotating BTZ black hole, with
$2\pi r_+$ the horizon perimeter.  In the fluctuation
analysis of the next two subsections we work to leading order in $c$,
where the distinction between $c$ and $c-1$ plays no role.

\subsection{Gaussian Fluctuations Around the Saddle}
\label{sec:fluctuations}

The saddle point of Section~\ref{sec:saddle} gives the leading
thermodynamics.  Because we have an explicit statistical description, we can go further and compute the
fluctuations of the ensemble.  Writing $P=P_*+\delta P$ and expanding
\eqref{eq:f-def},
\beq\label{eq:f-expand}
  f(P) \;=\; f(P_*) \;+\; \frac12\,f''(P_*)\,(\delta P)^2
  \;+\; \mathcal{O}\bigl((\delta P)^3\bigr)\, ,
  \qquad
  f''(P_*) \;=\;
  \frac{\de^2}{\de P^2}\log S_{0P}\bigg|_{P_*} \;-\; 2\beta_L\, ,
\eeq
where we used $4\pi i\tau=-2\beta_L$.  Differentiating
\eqref{eq:dlog-S} once more,
\beq\label{eq:d2log-S}
  \frac{\de^2}{\de P^2}\log S_{0P}
  \;=\;
  -\frac{(2\pi b)^2}{\sinh^2\bigl(2\pi bP\bigr)}
  \;-\;
  \frac{(2\pi/b)^2}{\sinh^2\bigl(2\pi P/b\bigr)}\, .
\eeq
The arguments at the saddle are
$\pi\lambda_L\bigl(1+O(1/k)\bigr)$ and $2\pi s_*$, as in
\eqref{eq:saddle-arguments}.  The second term is therefore
exponentially small in $s_*$, and the first is $O(b^2)=O(1/k)$, so
both are negligible against $2\beta_L$ and
\beq\label{eq:fpp-result}
  -f''(P_*) \;=\; 2\beta_L\,\bigl(1+O(1/k)\bigr)\, .
\eeq
The Gaussian integral around the saddle is then
\beq\label{eq:Gaussian-result}
  I \;\approx\; e^{f(P_*)}\,\sqrt{\frac{2\pi}{-f''(P_*)}}
  \;=\; e^{f(P_*)}\,\sqrt{\frac{\pi}{\beta_L}}\, ,
\eeq
and the spread of the Liouville momentum about the saddle is
\beq\label{eq:VarP}
  \bra(\delta P)^2\ket \;=\; \frac{1}{-f''(P_*)}
  \;=\; \frac{1}{2\beta_L}\, .
\eeq

Since $h_P=\frac{c-1}{24}+P^2$ by \eqref{eq:h-P}, the additive
constant drops out of the variance and the fluctuation of the
conformal weight is the fluctuation of $P^2$.  With
$P^2-P_*^2 = 2P_*\,\delta P+(\delta P)^2$, the first term dominates
the second by a factor of order $\sqrt{k}$, so
\beq\label{eq:Varh}
  \text{Var}(h) \;=\; 4\,P_*^2\;\bra(\delta P)^2\ket
  \;=\; \frac{2P_*^2}{\beta_L}
  \;=\; \frac{\pi^2 c}{3\,\beta_L^{3}}\, ,
\eeq
using $P_*=\pi Q/\beta_L$ from \eqref{eq:on-shell-eval} and $c-1=6Q^2$,
so that $P_*^2=\pi^2(c-1)/(6\beta_L^2)$, which is $\pi^2c/(6\beta_L^2)$
at leading order in $1/k$.  The antiholomorphic sector is independent,
and the same steps give $2\bar P_*^2/\beta_R=\pi^2c/(3\beta_R^3)$ for
the corresponding variance there.

It is worth noting that \eqref{eq:Varh} also follows without expanding
around the saddle.  The spectral integral \eqref{eq:I-def} is a
Boltzmann sum with $\beta_L$ conjugate to $P^2$, so the cumulants of
$P^2$ are derivatives of $\log I$ with respect to $\beta_L$,
\beq\label{eq:cumulants}
  \bra P^2\ket \;=\; -\frac{\de}{\de\beta_L}\,\log I\, ,
  \qquad
  \text{Var}(h) \;=\; \frac{\de^2}{\de\beta_L^2}\,\log I\, ,
\eeq
and inserting $\log I\approx\pi^2(c-1)/(6\beta_L)$ from
\eqref{eq:on-shell-eval} gives $\text{Var}(h)=\pi^2(c-1)/(3\beta_L^3)$,
which is \eqref{eq:Varh} at leading order in $1/k$.  Had we instead used
$\log(I/\eta)$, the same steps would return $\text{Var}(L_0)$, with the
descendants restoring $c-1\to c$.

\subsection{Fluctuations of the Entropy and the Horizon Area}
\label{sec:area-fluct}

To turn \eqref{eq:Varh} into a statement about the entropy we need the
response of $S_L$ to a change in $h$.  From the Cardy form
$S_L=2\pi\sqrt{c\,h/6}$ of \eqref{eq:S-BH-chiral},
\beq\label{eq:dS-dh}
  \frac{\de S_L}{\de h}\bigg|_{h_*}
  \;=\; \frac{S_L}{2h_*} \;=\; \beta_L\, ,
\eeq
which is the first law for the chiral sector.  The fluctuations are
Gaussian and small, so linear response is enough, and
\beq\label{eq:VarS-chiral}
  \text{Var}(S_L)
  \;=\;
  \left(\frac{\de S_L}{\de h}\right)^{\!2}\text{Var}(h)
  \;=\;
  \beta_L^2\;\frac{\pi^2c}{3\beta_L^{3}}
  \;=\;
  \frac{\pi^2 c}{3\,\beta_L}
  \;=\; S_L\, ,
\eeq
the last step by \eqref{eq:S-BH-chiral}.  The right-moving sector
gives $\text{Var}(S_R)=S_R$ in the same way.  The two chiral integrals
in \eqref{eq:Z-BTZ-full} factorise, so the sectors fluctuate
independently and their variances add,
\beq\label{eq:VarS-total}
  \text{Var}(S) \;=\; \text{Var}(S_L)+\text{Var}(S_R)
  \;=\; S_L+S_R \;=\; S\, .
\eeq
The variance of the Bekenstein-Hawking entropy is the entropy itself.
This is a Poisson-like relation, and it holds for the rotating black
hole as it stands, with $S$ the total entropy \eqref{eq:S-BH-total}.

The relation is not an accident of the $S$-kernel.  Each chiral
integral is a Boltzmann sum with $\beta_L$ conjugate to $h$, so
$\text{Var}(S_L)=\beta_L^2\,\text{Var}(h)$ is the heat capacity of that
sector, and a free energy of the Cardy form $\log I\propto1/\beta_L$
has $C_L=S_L$.  The two sectors combine into the familiar
thermodynamic statement: with $E=h_L+h_R$ and $J=h_L-h_R$ as in
\eqref{eq:BTZ-L0-main}, and using $\beta_{L,R}=\beta(1\mp\Omega_H)$
from \eqref{eq:smoothness-metric},
\beq\label{eq:first-law-rotating}
  \delta S \;=\; \beta_L\,\delta h_L \;+\; \beta_R\,\delta h_R
  \;=\; \beta\,\bigl(\delta E - \Omega_H\,\delta J\bigr)\, ,
\eeq
which is the first law of the rotating black hole, and
\eqref{eq:VarS-total} is the statement that the heat capacity at fixed
angular velocity equals the entropy.  What the computation above
supplies is the microscopic content: the fluctuating variable is the
Liouville momentum, which is the label of the Wilson-line microstate
and therefore of the holonomy.

Because the entropy is the horizon perimeter in Planck units, this
translates immediately into a statement about the area.  With
$S=A/4G_N$ and $A=2\pi r_+$,
\beq\label{eq:VarA}
  \text{Var}(A) \;=\; (4G_N)^2\,\text{Var}(S) \;=\; 4\,G_N\,A\, ,
  \qquad
  \frac{\Delta A}{A} \;=\; \sqrt{\frac{4G_N}{A}} \;=\; \frac{1}{\sqrt S}\, .
\eeq
The same answer can be read off directly from the holonomy, which
makes the origin of the fluctuation explicit.
At the leading order in $1/k$ used here, by
\eqref{eq:saddle-arguments} the horizon radius
$r_+=\frac12(\lambda_L+\lambda_R)=b\,(P+\bar P)$ is a linear function
of the two Liouville momenta, so \eqref{eq:VarP} gives
\beq\label{eq:VarA-direct}
  \text{Var}(A) \;=\; 4\pi^2b^2
  \Bigl[\bra(\delta P)^2\ket + \bra(\delta\bar P)^2\ket\Bigr]
  \;=\; 2\pi^2 b^2\left(\frac{1}{\beta_L}+\frac{1}{\beta_R}\right),
\eeq
and since $A=2\pi^2(\beta_L^{-1}+\beta_R^{-1})$ by
\eqref{eq:lambda-from-saddle}, this is $4G_NA$ once $b^2\approx1/k$
and $4G_N=1/k$ are used.  The area fluctuation is the spread of the
boost parameter of the spatial holonomy in the ensemble.

At the same leading order, the inner horizon is
$r_-=\frac12(\lambda_L-\lambda_R)=b(P-\bar P)$,
so the same two variances give the full covariance matrix of the
radii,
\beq\label{eq:Var-rpm}
  \text{Var}(r_\pm) \;=\; \frac{b^2}{2}
  \left(\frac{1}{\beta_L}+\frac{1}{\beta_R}\right),
  \qquad
  \text{Cov}(r_+,r_-) \;=\; \frac{b^2}{2}
  \left(\frac{1}{\beta_L}-\frac{1}{\beta_R}\right).
\eeq
The covariance vanishes at $\theta=0$.  The non-rotating ensemble
has $\bra J\ket=0$ while $J$ still fluctuates, and it does so
independently of the entropy.  To Gaussian order, $\text{Var}(S)=S$ is
then unchanged if one further restricts to $J=0$, so the non-rotating
results of this section are limits of the rotating ones at the level
of the fluctuations and not only at the level of the saddle.

The fractional fluctuation $\Delta A/A=1/\sqrt S$ is small for a
macroscopic black hole and becomes of order one when $S$ does.  Since
$S=2\pi k\,r_+$, the relative variance is
$\text{Var}(S)/S^2=1/S\sim1/(k\,r_+)$, so for $r_+$ of order one the
fluctuations are a one-loop effect in the Chern-Simons expansion, as
one would expect of a quantum correction\footnote{There is no tension with the fluctuations being ``thermal": because $S\sim k$ plays the role of $1/\hbar$, the thermal spread of the ensemble is automatically of one-loop size.
  Equivalently, in the purification \eqref{eq:TFD} of
  Section~\ref{sec:modular-ensemble}, the same variance is the
  quantum-mechanical variance of a one-sided observable in the pure
  state $|{\rm TFD}\rangle$.}.

\medskip\noindent\textbf{Comparison with semi-classical gravity.}\quad
Relations of the form $\text{Var}(A)\propto G_N A$ have been obtained
recently by quite different routes.  Parikh and
Pereira~\cite{Parikh:2024area} computed the renormalised graviton
propagator in linearised quantum gravity on the four-dimensional
Schwarzschild background in the Hartle-Hawking state, and found
$\Delta A\sim r_H\,l_P$, with $r_H$ the Schwarzschild radius and
$l_P=\sqrt{G\hbar}$ the Planck length, which is
$\text{Var}(A)\sim G\,A$.  Ciambelli, He and
Zurek~\cite{Ciambelli:2025area} obtained a lower bound of the same
parametric form for a finite causal diamond in $(d+2)$-dimensional
Minkowski spacetime, from the covariant phase space of a stretched
horizon.

These are not calculations of the same observable, and the $O(1)$
coefficients are not directly comparable.  But they are suggestive. The graviton computation
holds the coordinate sphere $r=r_H$ fixed and lets its shape
fluctuate, so that the horizon radius does not itself fluctuate, while
in the ensemble above the area varies because the boost parameter of
the spatial holonomy, and with it $r_+$, varies across the
microstates. Ours is moreover a variance at fixed $\beta_L$ and
$\beta_R$, that is, at fixed temperature and angular potential.  What the present setting supplies is a statistical-mechanical reading of
the area fluctuation.  The fluctuating variable is the label of an
explicit Wilson-line microstate, and the variance is a spread over the
same ensemble whose counting gives the entropy.  Parikh and Pereira
compute the variance of the area operator in a single fixed state, and
observe that its relation to fluctuations of the black hole entropy
remains unclear, since in the absence of a non-perturbative holographic
description the area does not carry a statistical-mechanical
interpretation as entropy~\cite{Parikh:2024area}. Note also that 4D Schwarzschild, with its negative specific heat, is a transient object and is therefore not the easiest place to formulate these statements.

\subsection{Thermal AdS$_3$: A Wilson Tower of Conical Defects}

Our discussion so far naturally involved only states above the BTZ threshold, which is loosely the same as being in the Cardy regime: we placed the microstates in the (grand-) canonical ensemble and noticed that the result has a geometric interpretation in terms of smoothness of the saddle configuration. It is natural to wonder if there are any statements one can make regarding microstates below the BTZ threshold. This is not the regime that is under the jurisdiction of Cardy, so one might think that there is nothing much we can say about it. But while the kinematics of modular invariance does not seem directly instructive here, dynamical statements about the actual spectrum of the CFT can teach us a few things. 

The key dynamical statement is that the spectrum of a holographic CFT seems often to be organized in terms of local (generalized free) fields in the bulk, rather than in terms of a structureless spectrum of isolated conical defects. In large-$N$ gauge theories, this is of course naturally realized in terms of the single- and multi-trace sectors. It was recently noted in \cite{Vishal}, building on the works of \cite{Spool1, Spool2, GMY}, that a bulk local field is naturally viewed as a tower of conical defect Wilson lines organized in terms of a seed primary and its associated multi-trace composites. This picture provides a bulk interpretation for the one loop determinant of the thermal gas of ``gravitons" (more generally, light fields) one expects below the Hawking-Page transition: it is simply the sum of thermal traces of each primary in the multi-trace tower of Wilson lines \cite{Vishal}.

\section*{Acknowledgments}

We thank Apoorva Asthana, ChatGPT/Claude, Vishal Gayari and Pradipta Pathak for discussions. This work is a natural follow-up of the ideas in \cite{CKPSP}. Some (though not all) of the conclusions here have already been reported earlier, scattered across the discussions in \cite{CKPSP, Vishal,PradiptaVishal}. A key technical ingredient here is that we have made things more explicit in terms of quantum Drinfel'd-Sokolov reduction. We have also provided the details of some of the derivations, which were previously simply mentioned as results.

\appendix

\section{States from Path Integrals: The 
  $SU(2)_k$ Prototype}\label{sec:compact-prototype}

This appendix gives a self-contained, pedagogical account of the
path-integral state construction used in the main text, but in the
compact setting of $SU(2)_k$ Chern-Simons theory. This is in principle entirely well-known, but we try to emphasize points that we could not easily decipher from the literature. Two questions are addressed.  First,
what does a Euclidean path integral on a half-infinite geometry
prepare when the bulk Hamiltonian vanishes, as it does in a
topological theory?  The familiar slogan -- ``Euclidean evolution
projects onto the lowest-energy state'' -- is a statement about
gapped systems, and it needs to be replaced suitably.
Second, what exactly does the semi-infinite Wilson line prepare on the
punctured disc?  We will see that the invariant answer is a
\emph{multiplet}-valued affine primary: the affine-primary condition
$J^a_{n>0}\Psi=0$ is a theorem, while the selection of a particular
vector (e.g.\ the highest-weight state) is a choice of boundary data
at the far endpoint of the Wilson line.  This sharpened statement is
what transfers to the noncompact gravitational problem, and it is what
makes the continuous-series construction of
Section~\ref{sec:super-threshold} -- where no extremal vector exists
-- structurally identical to the discrete-series one.

The gravitational $SL(2,\IR)$ problem of the main text differs from
this prototype in some ways, especially because the physical
Brown-Henneaux Virasoro algebra is obtained only after
Drinfel'd-Sokolov reduction.  The compact example should therefore be
read as a prototype for the punctured-disc Hilbert space and for the
state-preparation logic, not as a literal model of the full
gravitational reduction.

\subsection{Half-line Path Integrals in Quantum Mechanics}\label{app:QM-warmup}

Consider first ordinary quantum mechanics with Hamiltonian $H$.  The
Euclidean path integral on $\tau\in[-T,0]$, with the
boundary value $q(0)=q$ fixed and with initial data described by a
state $|\chi\ket$ at $\tau=-T$, computes
\beq\label{eq:QM-halfline}
  \Psi_T(q)
  \;=\;
  \bra q|\,e^{-TH}\,|\chi\ket\, .
\eeq
If $H$ has a unique ground state $\psi_0$ separated by a gap, then as
$T\to\infty$
\beq\label{eq:QM-projection}
  \Psi_T(q)
  \;\longrightarrow\;
  e^{-TE_0}\,\psi_0(q)\,\bra \psi_0|\chi\ket\, ,
\eeq
so that, up to normalization, the prepared state is $\psi_0$
regardless of $|\chi\ket$.  This is the textbook statement that the
half-line path integral ``prepares the ground state.''  It is worth
being precise about what this means: the far-end data $|\chi\ket$ never
disappears from the formula. It is merely damped into an overall
constant.  Two pieces of small print matter for
us:
\begin{enumerate}
\item If the minimal-energy subspace is \emph{degenerate}, the
  projection lands in that subspace, but the direction \emph{within}
  it remains entirely determined by $|\chi\ket$ -- i.e., by the
  boundary data at the far end.
\item If $H=0$, then $e^{-TH}=\mathbf{1}$ and
  \eqref{eq:QM-halfline} is just the overlap $\bra q|\chi\ket$.
  Nothing is projected: the state prepared \emph{is} the far-end data,
  transported to the cut. This is the case in topological theories of interest to us.
\end{enumerate}
One might hope to evade the choice of $|\chi\ket$ by ``specifying
nothing'' at $\tau=-T$.  But operationally, specifying nothing can
only mean integrating over the far endpoint $q(-T)=q'$ with some
measure $\mu(q')$, and
\beq\label{eq:QM-nothing}
  \int dq'\,\mu(q')\,\bra q|e^{-TH}|q'\ket
  \;=\;
  \begin{cases}
    \;\psi_0(q)\displaystyle\int dq'\,\mu(q')\,\psi_0^*(q') \cdot e^{-TE_0}
    & \hbox{(gapped, $T\to\infty$)}\,,\\[10pt]
    \;\mu(q)
    & \hbox{($H=0$)}\, .
  \end{cases}
\eeq
In the gapped case the $\mu$-dependence collapses into normalization.  In
the $H=0$ case the wavefunction one prepares \emph{is} the measure one
chose at infinity.
The same conclusion holds for first-order (phase-space) systems: for
$S=\int d\tau\,(-i\,p\,\dot q+H)$ with $H=0$, the kernel over any
Euclidean interval is $\bra q|q'\ket=\delta(q-q')$, and the half-line
$p$--$q$ path integral returns the input wavefunction unchanged.

The doctrine that survives all cases is this: \emph{a path integral on
a manifold with boundary defines a wavefunctional of the boundary
data, and the state it represents is fixed by (i) the insertions and
(ii) the conditions supplied at the other ends of the manifold.}   

\subsection{The Punctured-disc Hilbert Space}\label{app:punctured-disc}

Now consider $SU(2)_k$ Chern-Simons theory on
$D^2\times\IR_\tau$, with a Wilson line in the spin-$j$
representation $V_j$ inserted at the origin of the disc and running
along the Euclidean time direction.  At fixed $\tau$, the spatial
slice is a disc with one puncture, and the puncture carries the label
$j$.

The CS/WZW correspondence states that the Hilbert space of
Chern-Simons theory on a surface with boundary is described by the
corresponding chiral WZW conformal blocks on that surface
\cite{WittenJones,EMSS}.  In the present case, the boundary of the
disc is a circle carrying the chiral $SU(2)_k$ WZW degrees of freedom,
and the puncture selects the affine sector.  Thus
\beq\label{eq:H-SU2}
  {\cal H}(D^2;j)
  \;\cong\;
  \widehat V_j\, ,
\eeq
where $\widehat V_j$ is the integrable highest-weight module of
$\widehat{\mathfrak{su}}(2)_k$ built on the finite-dimensional
spin-$j$ representation $V_j$.

In radial quantization of the boundary WZW model, this statement is
the state--operator correspondence \cite{DMS}.  The WZW primary
$\Phi_j$ associated with the puncture is a $V_j$-valued field: its
components $\Phi_{j,m}$, $m=-j,\ldots,j$, inserted at the origin
create the degenerate multiplet of affine primary states $|j,m\ket$,
and the negative current modes generate the rest of the affine module:
\[
  J^{a_1}_{-n_1}\cdots J^{a_p}_{-n_p}
  |j,m\ket\,,
  \qquad n_i>0\, .
\]
For compact $SU(2)_k$, integrability restricts the allowed labels to
\beq
  j \in
  \left\{0,\frac12,1,\ldots,\frac{k}{2}\right\}.
\eeq
Thus there is a finite set of allowed puncture labels, and each label
selects a corresponding affine module.  Note that \eqref{eq:H-SU2} is
an isomorphism of Hilbert spaces, i.e.\ a statement about
\emph{which} states exist on the punctured disc.  It does not by
itself say which vector a given path integral prepares. That is the
question we turn to next.

\subsection{The Half-cylinder State: Endpoint Data and
  the Primary}\label{app:half-cylinder-state}

Perform the Euclidean path integral on the half-cylinder
\beq
  M_- \;=\; D^2\times(-\infty,0]\, ,
\eeq
with the Wilson line running from $\tau=-\infty$ to $\tau=0$ along the
core $\{0\}\times(-\infty,0]$.  As mentioned already, 
an \emph{open} Wilson line is not a number: it is the path-ordered
exponential ${\cal P}\exp\int A$, a $V_j$-endomorphism-valued
functional of the connection, with a free representation index at each
end.  The index at $\tau=0$ is part of the definition of the states on
the punctured disc.  The index at $\tau=-\infty$, however, is free:
under a gauge transformation it rotates in $V_j$, so a well-defined
(gauge-covariantly meaningful) state is obtained only after it is
contracted with a vector $v\in V_j$.  The half-cylinder path integral
is therefore, naturally viewed as  a \emph{linear map}
\beq\label{eq:PI-map-SU2}
  \Psi_j:\;V_j\;\longrightarrow\;{\cal H}(D^2;j)\,,
  \ \
  v\;\longmapsto\;|\Psi_j(v)\ket
  \;=\;
  \int [\mathcal D A]\,
  W_j\bigl(\{0\}\times(-\infty,0]\,;\,v\bigr)\,
  e^{-S^E_{\rm CS}[A]}\,
  \bigl|A\bigr|_{\tau=0}\ket\, ,
\eeq
where $W_j(\gamma;v)$ denotes the open Wilson line with its far
endpoint contracted with $v$\footnote{One might again try to dodge the choice
of $v$ by averaging suitably. But averaging the endpoint over the group
with the invariant (Haar) measure projects $v$ onto the invariant
subspace ${\rm Inv}(V_j)$, which vanishes for every $j\neq 0$.  For a
nontrivial representation, ``democratically summing over the far-end
data'' does not produce a natural state -- it produces zero. A charged
endpoint cannot be gauge-invariantly hidden. }.   This is the
Chern-Simons incarnation of the $H=0$ lesson of
Appendix~\ref{app:QM-warmup}.

Although the bulk
theory is topological -- the bulk Hamiltonian is a constraint, so
there is no bulk analogue of the damping in
\eqref{eq:QM-projection} -- the \emph{edge} theory is not.  Once the
boundary polarization is fixed, the boundary circle carries the chiral
WZW model, and Euclidean translation along the semi-infinite boundary
cylinder is generated by the edge Sugawara operator $L_0^{\rm Sug}$, whose
eigenvalue on the affine primaries of the spin-$j$ sector is
\beq\label{eq:sugawara-weight}
  h_j^{SU(2)}
  \;=\;
  \frac{j(j+1)}{k+2}\, ,
\eeq
and $h_j^{SU(2)}+n$ on level-$n$ current descendants.  The factor
$e^{-T L_0^{\rm Sug}}$, $T\to\infty$, therefore genuinely damps: it removes
every current descendant relative to the primaries.  But it can go no
further.  $L_0^{\rm Sug}$ commutes with the zero modes $J^a_0$, so all $2j+1$
affine primaries $|j,m\ket$ are \emph{exactly degenerate} at
\eqref{eq:sugawara-weight}, and the Euclidean damping is blind to the
$m$-direction.  We are squarely in case~1 of
Appendix~\ref{app:QM-warmup}: the projection lands on the degenerate
primary multiplet, and the vector within it is fixed by the far-end
data -- here, the endpoint contraction $v$.

Putting the two observations together yields a uniqueness statement:
the map \eqref{eq:PI-map-SU2} is completely fixed by symmetry, up to a
single overall constant.  The argument has three steps.

\emph{Step 1: the image lies in the affine-primary subspace of
$\widehat V_j$.}  This is the content of the damping argument just
given: whatever $v$ is fed in, the output state has components only
along the span of the $2j+1$ primaries $|j,m\ket$ annihilated by all
$J^a_{n>0}$.  (The same statement will be re-derived, using boundary
CFT methods and with no appeal to energetics, in
Appendix~\ref{app:contour-argument}.)

\emph{Step 2: the map intertwines the zero-mode $\mathfrak{su}(2)$
action.}  By this we mean that rotating the input vector and then
performing the path integral gives the same answer as performing the
path integral and then rotating the output state with the boundary
charges.  To see this, consider a rigid (constant) gauge
transformation $g\in SU(2)$ on the half-cylinder.  Constant
transformations are compatible with the boundary polarization: they
survive as the global symmetry of the edge theory, implemented on the
edge Hilbert space by the charges $J^a_0$.  The Chern-Simons action
and the measure are invariant, so the path integral is unchanged if
one simultaneously transforms everything that carries a gauge index.
There are exactly two such objects: the free Wilson-line endpoint at
$\tau=-\infty$, where $v\to g\cdot v$ in $V_j$, and the boundary data
at $\tau=0$, on which $g$ acts through the global charge.  Invariance
under doing \emph{both} equates the two actions on the output:
\beq\label{eq:intertwining-SU2}
  |\Psi_j(g\cdot v)\ket
  \;=\;
  U(g)\,|\Psi_j(v)\ket\, ,
  \qquad
  U(g)=e^{\,i\theta_a J^a_0}\, .
\eeq
This is nothing but the Ward identity for global gauge rotations.
(Looking ahead slightly: a local avatar of this appears in Appendix~\ref{app:contour-argument}. The
$n=0$ contour integral picks up the simple pole at the endpoint
insertion and acts as the generator $t^a$ on the dangling index --
the zero modes measure the endpoint charge.)

\emph{Step 3: Schur's lemma.}  On the affine-primary subspace the zero
modes $J^a_0$ act exactly as the spin-$j$ matrices $t^a$ act on $V_j$:
the grade-zero floor of the module is the horizontal
representation.  So, by Steps 1 and 2, the path integral is a linear
map between two irreducible copies of $V_j$ that commutes with the
$\mathfrak{su}(2)$ action, and Schur's lemma forces it to be a
multiple of the identity.%
\footnote{Step 1, where the restriction to the primary subspace happens, is the important step.  The full module $\widehat V_j$ contains many further
spin-$j$ copies at higher grades -- the current descendants also
organize into $\mathfrak{su}(2)$ multiplets -- so an intertwiner
$V_j\to\widehat V_j$ is \emph{not} unique by itself.  Uniqueness holds
because the geometry independently forces the image onto grade zero.}

The map is therefore canonical: it comes with no choices other than which vector $v$ is fed into it.  This gives the
invariant, choice-free statement of what the path integral constructs:
\begin{quote}
\emph{The undecorated semi-infinite Wilson-line path integral is the
canonical embedding $V_j\hookrightarrow \widehat V_j$ of the zero-mode
multiplet onto the affine-primary subspace, unique up to
normalization.  A state is obtained only after contracting the free
endpoint index: the contraction $v=|j,m\ket$ prepares the affine
primary $|j,m\ket$.}
\end{quote}
In particular, choosing the highest-weight vector
$v=|j,\mathrm{hw}\ket$ gives
\beq\label{eq:hw-state-SU2}
  |\Psi_j\ket
  \;=\;
  |j;\mathrm{hw}\ket
  \;\in\;
  \widehat V_j\, ,
\eeq
which satisfies
\beq\label{eq:hw-conditions}
  J^a_n\,|j;\mathrm{hw}\ket=0
  \quad \forall\, n>0\,,
  \qquad
  J^+_0\,|j;\mathrm{hw}\ket=0\,,
  \qquad
  J^3_0\,|j;\mathrm{hw}\ket
  =j\,|j;\mathrm{hw}\ket\, .
\eeq
The logical status of these conditions is not uniform, and keeping
them apart is the main point of this appendix.  The first condition
-- the affine-primary condition -- is automatic: it holds for every
choice of $v$.  The remaining two express the \emph{choice}
$v=|j,\mathrm{hw}\ket$.  For $SU(2)$ this is a natural choice, since
the highest-weight vector is distinguished within the unitary
multiplet. But it is a prescription supplied at $\tau=-\infty$, not something that came from dynamics.  Any presentation of the half-cylinder
path integral that yields a specific vector without mentioning a
choice has smuggled the contraction in somewhere, e.g.\ through a
worldline gauge-fixing at $\tau\to-\infty$ or through an
$i\epsilon$-type regulator that damps onto the extremal weight.  These
are legitimate prescriptions, they are not outputs.

\subsection{Regularity at the Puncture: Contour
  Argument}\label{app:contour-argument}

The affine-primary condition can be derived in a way that makes its kinematic character manifest.
Map the semi-infinite boundary cylinder to the punctured unit disc by
\beq
  z \;=\; e^{\tau+i\phi}\, ,
\eeq
so that the $\tau=0$ boundary circle becomes $|z|=1$ and the far end
$\tau=-\infty$ becomes the origin $z=0$.  In this frame the entire
representation content of the Wilson line is concentrated at $z=0$ as
the $V_j$-valued primary insertion $\Phi_j(0)$ of
Appendix~\ref{app:punctured-disc}, and the state on $|z|=1$ is the
disc one-point conformal block with that insertion.  Acting with a
current mode on the state means inserting the contour integral
\beq
  J^a_n \;=\; \oint \frac{dz}{2\pi i}\; z^{n}\, J^a(z)
\eeq
on a circle just inside the boundary.  Away from insertions the
current is holomorphic -- this is the chiral Ward identity, and in
bulk language it is Gauss's law: flatness ($F=0$) allows the circle on
which the charge is measured to be deformed freely.  The contour can
therefore be slid inward until it hits the puncture at $z=0$.

What does the puncture look like to the current?  A \emph{bare}
Wilson-line endpoint is a pointlike source of charge in the
representation $V_j$.  The most singular thing the
current can do near it is measure that charge:
\beq\label{eq:endpoint-OPE}
  J^a(z)\,\Phi_j(0)
  \;\sim\;
  \frac{\bigl(t^a\,\Phi_j\bigr)(0)}{z}
  \;+\;\hbox{regular}\, ,
\eeq
a simple pole whose residue is the spin-$j$ generator $t^a$ acting on
the dangling endpoint index.  Higher-order poles would signal
additional structure attached to the endpoint -- derivative
dressings, i.e.\ current descendants stuck to the tip.  ``Undressed
endpoint $\Rightarrow$ at most a simple pole'' is the precise content
of ``the state is an affine primary,'' and it is a statement about
what was inserted.  Evaluating the moments,
\beq
  J^a_n\,\Psi
  \;\longleftrightarrow\;
  \oint \frac{dz}{2\pi i}\,
  \Bigl[\, z^{\,n-1}\,\bigl(t^a\Phi_j\bigr)(0) + O(z^{n})\,\Bigr]\, ,
\eeq
one reads off:
\begin{itemize}
\item $n\geq 1$: the integrand is holomorphic at the origin
  ($z^{\,n-1}$ with $n-1\geq0$), so there is no residue and
  $J^a_{n\geq1}\Psi=0$;
\item $n=0$: the residue is $\bigl(t^a\Phi_j\bigr)(0)$ -- the zero
  modes act as the ordinary finite-dimensional generators on the
  endpoint index, rotating the multiplet;
\item $n\leq-1$: the integrand has genuine higher-order poles; these
  modes do not vanish, and create the current descendants.
\end{itemize}
A loose physical picture is: the modes $J^a_n$ are the
multipole moments of the edge charge distribution, weighted by
$z^{n}$.  The Wilson line presents the edge with a point charge
sitting exactly at the origin, and every positive moment of a point
charge at the origin vanishes.  The zeroth moment is the total charge, which
is exactly the $\mathfrak{su}(2)$ action on the multiplet.  

\subsection{Affine Tower and Lessons for Gravity}\label{app:affine-tower}

The full affine module contains the current-algebra descendants
\beq\label{eq:descendants-SU2}
  J^{a_1}_{-n_1}\,
  J^{a_2}_{-n_2}\cdots
  J^{a_p}_{-n_p}\,
  |j,m\ket\,,
  \qquad n_i>0\, ,
\eeq
with Sugawara weight $h_j^{SU(2)}+\sum_i n_i$.  These are the states
preparable by \emph{decorated} half-cylinders -- inserting boundary
current modes at $\tau<0$.  The module
$\widehat V_j$ of \eqref{eq:H-SU2} is the answer to the question
``what states exist on the punctured disc". The reachable set of the
\emph{undecorated} path integral \eqref{eq:PI-map-SU2} on the other hand, is only the
degenerate horizontal multiplet $V_j\subset\widehat V_j$.  All of
these states share the same puncture label $j$. Using this
compact example as an analogy for the gravitational problem, this
fixed label is analogous to fixing the primary holonomy data.

We conclude with some comments on the gravitational case.  First, Sugawara gives a Virasoro
action on the affine module, but it does not identify the affine
module one-to-one with the Virasoro descendant module of a single
primary. The affine module is generally larger.  In the gravitational
$SL(2,\IR)$ problem one must in addition impose the Drinfel'd-Sokolov
reduction associated with the Brown-Henneaux boundary conditions, and
only after that reduction should the surviving physical sector be
interpreted as a Virasoro primary together with its Virasoro
descendants.

Second -- the
half-cylinder canonically prepares the \emph{affine-primary multiplet}
(as the intertwiner $V_j\hookrightarrow\widehat V_j$), and a
particular vector only after an endpoint contraction is chosen.  For
compact $SU(2)$ the unitary multiplet has a distinguished extremal
vector (the highest-weight state, annihilated by $J^+_0$ with
$J^3_0$-eigenvalue $+j$).  In the noncompact discrete series ${\cal D}^{+}_j$ of the
main text, there is again a distinguished extremal vector -- now
the lowest-weight state, annihilated by $J^-_0$ with the same
eigenvalue $+j$.  In the principal continuous series of Section~\ref{sec:super-threshold}, there is no extremal vector at all. But DS reduction ensures that none is needed. In every case the invariant content of the path integral is the multiplet-valued affine primary. In the gravitational problem the endpoint data are unreduced affine data, but they are redundant because of the reduction. The Whittaker/zero-mode analysis of Section~\ref{sec:zero-mode-conical} and Appendix~\ref{app:DS-reduction} shows that, at fixed Casimir, the BRST complex has a single level-zero class in the discrete and continuous cases alike.

\section{Principal-Series Zero Mode and BRST Cohomology}
\label{app:DS-reduction}

This appendix presents the only representation-dependent ingredient of
the principal-series reduction that is needed in the main text: the
level-zero Brown-Henneaux constraint selects a single primary class at
fixed Casimir.  The full positive-level BRST cohomology is deferred to
\cite{Future}.

The unreduced horizontal representation is the principal continuous
series $\mathcal C_s^\epsilon$, with
\beq\label{eq:principal-app-data}
  J^3_0|s,\epsilon;m\rangle
  =m|s,\epsilon;m\rangle,
  \qquad
  m\in\epsilon+\IZ,
  \qquad
  C_2=s^2+\frac14\, .
\eeq
It has no extremal vector.  In a standard weight basis one may write
\beq\label{eq:principal-lowering-app}
  J^-_0|s,\epsilon;m\rangle
  =B_m|s,\epsilon;m-1\rangle,
\eeq
where $B_m\neq0$ for every allowed $m$ when $s>0$.  The precise
normalization of $B_m$ will not be needed.

As explained in Section~\ref{sec:DS-conical}, Brown-Henneaux boundary
conditions impose a nonzero constant on the boundary cylinder.
Keeping its normalization temporarily as $t\neq0$, the level-zero part
of the flowed BRST constraint is
\beq\label{eq:principal-whittaker-app}
  (J^-_0-t)|\psi\rangle=0\, .
\eeq
The reduction is naturally formulated in the formal direct-product
completion which can be written as $\prod_{m\in\epsilon+\IZ}\IC\,|s,\epsilon;m\rangle$ of the
zero-mode direction (with no growth condition)\footnote{For the principal series this completion is essential: the
Whittaker representative is a formal sum over the zero-mode basis and
is not normalizable in the unreduced principal-series norm.  But this is
not the normalizability condition on the physical state.  The physical
state is the reduced BRST cohomology class.  For real $P$ and $c>1$, we expect (and find evidence that) the cohomology is the Verma module $\mathcal V_{h_P}$. The latter has positive-definite
Shapovalov form, with $\langle h_P|h_P\rangle=1$.}, so we can write
\beq\label{eq:principal-formal-app}
  |\psi\rangle
  =
  \sum_{m\in\epsilon+\IZ}c_m|s,\epsilon;m\rangle\, .
\eeq
Substituting \eqref{eq:principal-lowering-app} gives the first-order
recursion
\beq\label{eq:principal-recursion-app}
  B_{m+1}c_{m+1}=t\,c_m\, .
\eeq
Since neither $t$ nor any $B_m$ vanishes, specifying one coefficient
fixes all the others in both directions.  Hence
\beq\label{eq:principal-kernel-app}
  \dim\ker(J^-_0-t)=1
\eeq
on the completed level-zero space.  The same recursion, with an
inhomogeneous term, shows that $J^-_0-t$ is surjective on this
completion.  With the ghost convention of Section~\ref{sec:zero-mode-conical},
this gives
\beq\label{eq:principal-H0-app}
  H^0_{\rm DS,flow}\big|_{N=0}\cong\IC,
  \qquad
  H^1_{\rm DS,flow}\big|_{N=0}=0\, .
\eeq
Thus the two-sided $J^3_0$ lattice of the unreduced principal series at fixed $s$ and fixed
$\epsilon$ produces one level-zero cohomology class.  In the main text
we set $\epsilon=0$ throughout. The Virasoro weight itself depends only
on the Casimir.

The conformal weight of this class does not have to be inferred from
the recursion.  The module-independent flowed Hamiltonian was derived
in Section~\ref{sec:microstate-tower},
\beq\label{eq:principal-L0-app}
  L_0^{\rm phys}
  =L_0^{\rm Sug}+L_0^{\rm gh}+\frac{k}{4}\, ,
\eeq
so at level zero
\beq\label{eq:principal-h-app}
  h
  =
  \frac{C_2}{k-2}+\frac{k}{4}
  =
  \frac{c-1}{24}+\frac{s^2}{k-2}
  =
  \frac{c-1}{24}+P^2,
  \qquad
  P=\frac{s}{\sqrt{k-2}}\, .
\eeq
The important point is that the $J^3_0$ dependence has already
cancelled in $L_0^{\rm phys}$.  The level-zero representative is spread
over the horizontal weight basis, but it nevertheless has a sharp
Virasoro weight fixed only by the Casimir.

Using Whittaker/BRST complex methods, we have explicitly checked that the BRST cohomology precisely reproduces the Verma module expectation at levels $N=0, 1, 2, 3, 4$ for all ghost numbers \cite{Future}. The direct calculation gets significantly more complicated at higher levels, compared to the trivial level-0 case we discussed above. We have also checked that $\dim(H^0|_N)=p(N)$, the number of partitions of $N$, at these levels. In other words, in all sectors examined, the cohomology is non-trivial only at ghost number zero, where its dimensionality matches the
Virasoro descendant count. What fails to adapt straightforwardly from the standard theorems on (unflowed) discrete series are the all-level proofs: but they fail irrespective of level, so the fact that we can show that it works by direct cohomology calculations at low levels can be viewed as indicating that it is the proof methods that are failing, and not the reduction itself. We hope to present a complete demonstration in \cite{Future}. We also note closely related earlier work by Dhillon \cite{Dhillon} and a very recent paper \cite{FFR}. These papers come from a more mathematical background, and we do not rule out the possibility that the result we want may already be established as a (corollary of a) formal theorem.

\end{document}